\documentclass[11pt,a4paper]{article} 
\usepackage{jheppub}
\pdfoutput=1

\usepackage{amsmath,amsfonts}
\usepackage{mathtools}
\usepackage{latexsym}
\usepackage{hhline}
\usepackage{stmaryrd}
\usepackage{color}
\usepackage{verbatim}
\usepackage{graphicx}
\usepackage{caption}
\usepackage{subcaption}
\usepackage{booktabs}
\usepackage{multirow}
\usepackage{epstopdf}
\usepackage[utf8]{inputenc}
\usepackage{slashed}
\usepackage{xcolor}
\usepackage{fancyvrb}
\usepackage{tikz}
\usepackage{tcolorbox}
\usepackage{hyperref}
\usepackage{cleveref}

\tcbuselibrary{skins}
\tcbuselibrary{listings}
\tcbuselibrary{breakable}

\def\eg{{\it e.g.}}

\newcommand{\be}{\begin{equation}}
\newcommand{\ee}{\end{equation}}
\def\bsp#1\esp{\begin{split}#1\end{split}}
\def\bpm{\begin{pmatrix}}
\def\epm{\end{pmatrix}}
\newcommand{\fr}{{\sc FeynRules}}
\newcommand{\hdmspectra}{{\sc HDMSpectra}}

\newcommand\aNLO{{\sc\small MadGraph5\_aMC@NLO}}

\newtcblisting{mgscript}[1][]{enhanced,listing only,
boxrule=0.4pt,left=3mm,right=2mm,top=1mm,bottom=1mm,
colback=white,
colframe=black,
boxed title style={size=small,colback=white!82!black,boxrule=0.4pt},
fonttitle=\ttfamily\footnotesize,coltitle=black,
sharp corners,rounded corners=southeast,arc is angular,arc=3mm,
underlay={
  \path[fill=white!70!black] ([yshift=3mm]interior.south east)--++(-0.4,-0.1)--++(0.1,-0.2);
  \path[draw=black,shorten <=-0.05mm,shorten >=-0.05mm] ([yshift=3mm]interior.south east)--++(-0.4,-0.1)--++(0.1,-0.2);
  },
drop fuzzy shadow,attach boxed title to top right={yshift=-2mm, xshift=-4mm},#1}

\preprint{CP3-23-71, CA21106} 

\title{Phenomenology of superheavy decaying dark matter from string theory}

\author[1]{Rouzbeh Allahverdi}
\author[2]{, Chiara Arina}
\author[5,6]{, Marco Chianese}
\author[3,4]{, Michele Cicoli}
\author[2,3,4]{, Fabio Maltoni}
\author[2,3,4]{, Daniele Massaro}
\author[7]{, Jacek K. Osi\'nski}

\affiliation[1]{Department of Physics and Astronomy, University of New Mexico, Albuquerque, NM 87131, USA}
\affiliation[2]{Centre for Cosmology, Particle Physics and Phenomenology (CP3), Universit\'e catholique de Louvain, B-1348 Louvain-la-Neuve, Belgium}
\affiliation[3]{Dipartimento di Fisica e Astronomia, Universit\`a di Bologna, via Irnerio 46, 40126 Bologna, Italy}
\affiliation[4]{INFN, Sezione di Bologna, viale Berti Pichat 6/2, 40127 Bologna, Italy}
\affiliation[5]{Dipartimento di Fisica ``Ettore Pancini'', Università degli studi di Napoli ``Federico II'', Complesso Univ. Monte S. Angelo, I-80126 Napoli, Italy}
\affiliation[6]{INFN, Sezione di Napoli, Complesso Univ. Monte S. Angelo, I-80126 Napoli, Italy}
\affiliation[7]{AstroCeNT, Nicolaus Copernicus Astronomical Center of the Polish Academy of Sciences, ul. Rektorska 4, 00-614 Warsaw, Poland}

\emailAdd{rouzbeh@unm.edu}
\emailAdd{chiara.arina@uclovain.be}
\emailAdd{michele.cicoli@unibo.it}
\emailAdd{marco.chianese@unina.it}
\emailAdd{fabio.maltoni@unibo.it}
\emailAdd{daniele.massaro5@unibo.it}
\emailAdd{josin@camk.edu.pl}

\abstract{We study the phenomenology of superheavy decaying dark matter with mass around $10^{10}$ GeV which can arise in the low-energy limit of string compactifications. Generic features of string theory setups (such as high scale supersymmetry breaking and epochs of early matter domination driven by string moduli) can accommodate superheavy dark matter with the correct relic abundance. In addition, stringy instantons induce tiny $R$-parity violating couplings which make dark matter unstable with a lifetime well above the age of the Universe. Adopting a model-independent approach, we compute the flux and spectrum of high-energy gamma rays and neutrinos from three-body decays of superheavy dark matter and constrain its mass-lifetime plane with current observations and future experiments. We show that these bounds have only a mild dependence on the exact nature of neutralino dark matter and its decay channels. Applying these constraints to an explicit string model sets an upper bound of ${\cal O}(0.1)$ on the string coupling, ensuring that the effective field theory is in the perturbative regime.}


\begin{document}

\maketitle
\flushbottom

\section{Introduction}

Despite various lines of evidence for the existence of dark matter (DM) in the Universe~\cite{Silk}, its nature remains an important unsolved problem at the interface of cosmology and particle physics. Weakly interacting massive particles (WIMPs) have been a promising candidate and have driven theoretical and experimental research in this area for a long time. The lack of any positive signals in direct, indirect, and collider searches has however led to the consideration of DM candidates much lighter or much heavier than the typical mass range for WIMP-like particles. The well-known bounds on the annihilation cross section imply that freeze-out in a standard thermal history results in DM overproduction for masses larger than 100 TeV~\cite{GK}, thereby requiring alternative scenarios for obtaining the correct relic abundance of superheavy DM with $m_{\rm DM}\gg 100$ TeV. 

Superheavy DM can easily evade direct detection experiments and collider searches. Standard indirect detection signals from DM annihilation are also highly suppressed in this case. This is because, in addition to small annihilation cross sections, such a signal scales like $n_{\chi,0}^2$ ($n_{\chi,0}$ being the current number density of DM particles). The situation will improve if DM is not absolutely stable as signals from decaying DM scale like $n_{\chi,0}$. This scenario is conceivable because symmetries that are typically invoked to make DM stable may be broken by high-energy effects (from, for example, string theory or quantum gravity). While this could provide an observational window to probe superheavy DM, a viable scenario must include lifetimes that are orders of magnitude longer than the age of the Universe~\cite{Chianese:2021htv, Chianese:2021jke, Esmaili:2015xpa, Arguelles:2022nbl, Murase:2012xs, Esmaili:2012us, Cohen:2016uyg, Kalashev:2016cre, HAWC:2017udy, Kachelriess:2018rty, Ishiwata:2019aet, Kalashev:2020hqc, Guepin:2021ljb, LHAASO:2022yxw, IceCube:2023ies, Fiorillo:2023clw}. This leads to another question on how the very tiny couplings required for such long lifetimes arise. 

In this paper, we address both questions regarding the origin of the abundance and indirect signals of superheavy DM in the context of an explicit construction within string theory. Successful inflation in this model results in a high scale of supersymmetry (SUSY) breaking (in the absence of fine tuning). The lightest SUSY particle (LSP) is stable by virtue of $R$-parity conservation, and is hence the DM candidate. The model gives rise to a non-standard cosmological history that involves one or more epochs of early matter domination (EMD) driven by string moduli. This can yield the observed relic abundance for a DM mass in the $10^{10}-10^{11}$ GeV range~\cite{Allahverdi:2020uax}. As we will see, this model can also accommodate decaying DM through tiny $R$-parity violating (RPV) interactions whose smallness is due to exponentially suppressed non-perturbative terms that are induced by instantons.

We find that, for setups which realise the Minimal SUSY Standard Model (MSSM), the resulting mass spectrum implies a Bino-like LSP. Small RPV effects lead to three-body decays of DM to Standard Model (SM) particles. We calculate the spectrum of high-energy gamma rays and neutrinos from these decays and constrain the DM mass-lifetime plane by imposing constraints from current experiments as well as sensitivity limits of future observations. These limits have a rather mild dependence on the exact RPV channel and do not change much even for a Higgsino-like or Wino-like LSP. The tightest bounds on the DM lifetime happen to be in the $10^{9}-10^{11}$ GeV mass range. As we will see, the resulting limits on the size of RPV couplings in the model imply an upper bound of ${\cal O}(0.1)$ on the string coupling, which ensures the effective field theory remains in the perturbative regime.          

This paper is organized as follows. In Sec.~\ref{sec:model} we will describe a string construction that can naturally realise a SUSY model with superheavy decaying DM. We then discuss how various RPV terms make the LSP unstable in Sec.~\ref{sec:spectrum}. In Sec.~\ref{sec:expconstr} we discuss the relevant current and future experiments that can constrain this type of DM by means of high-energy gamma rays and neutrinos. We describe the methodology used in the numerical computations in Sec.~\ref{sec:spectra}. In Sec.~\ref{sec:bounds_indep} we show for the first time the model-independent neutrino and gamma-ray constraints on superheavy DM decaying into three-body final states, as well as the sensitivity of current and future experiments to the string model presented here. We conclude the paper and give further outlook in Sec.~\ref{sec:concl}. Technical details on the renormalization group equations (RGEs) are given in App.~\ref{app:RGE}, while the case of superheavy sneutrino DM with two-body decays is described in App.~\ref{app:snudecay}.

\section{Superheavy dark matter from string theory}
\label{sec:model}

\subsection{General features of 4D string models}
\label{GenFeat}

Low-energy 4D string models have been explored in detail in recent decades, focusing in particular on the type IIB framework where moduli stabilisation is better under control. These studies revealed the following generic features:
\begin{itemize}
\item \textbf{High SUSY breaking scale:} Statistical studies of the distribution of the SUSY breaking scale in the string flux landscape have shown a (power-law or logarithmic) preference for higher scales of SUSY breaking \cite{Denef:2004cf, Broeckel:2020fdz}.\footnote{See however \cite{Cicoli:2022chj} which found a preference for lower scales of SUSY breaking when considering the zero cosmological constant limit of the joint distribution of the gravitino mass and the vacuum energy in models with antibrane uplifting.} 

\item \textbf{High inflationary scale:} Realising inflation and low-energy SUSY in string models is a difficult task \cite{Kallosh:2004yh}. This is indeed the case in most of the known string inflationary models \cite{Cicoli:2023opf}. The main obstacle comes from the requirement of matching the observed amplitude of the primordial density perturbations which fixes the inflationary energy scale at relatively high energies which, in turn, fixes also the value of the gravitino mass during inflation. In principle, the gravitino mass could evolve from the end of inflation to today \cite{Conlon:2008cj,Cicoli:2015wja}, or SUSY particles could be much lighter than the gravitino due to sequestering in the extra dimensions \cite{Blumenhagen:2009gk, Aparicio:2014wxa}, but these two features do not seem to be generic. On the other hand, high-scale inflation seems to imply also a high SUSY spectrum in general, regardless of the statistics in the string landscape.

\item \textbf{Moduli domination:} 4D string compactifications come along with a plethora of \emph{moduli} which are new scalar fields with gravitational couplings to matter. During inflation, these fields are naturally displaced from their minimum due to the inflationary energy density. Due to this initial misalignment, which is generically of order the Planck scale \cite{Cicoli:2016olq}, the moduli store energy when they start oscillating around their minima which happens when the Hubble scale $H$ becomes comparable to their mass. Being redshifted more slowly than radiation (since they behave as matter), the moduli quickly come to dominate the energy density of the Universe, giving rise to EMD epochs which can alter the standard thermal history from the end of inflation to Big-Bang Nucleosynthesis. In particular, the decay of the moduli can dilute any DM production which took place prior to their decay \cite{Acharya:2008bk, Allahverdi:2013noa, Allahverdi:2014ppa,Allahverdi:2020uax,Aparicio:2015sda, Aparicio:2016qqb, Cicoli:2022uqa}. 

\item \textbf{MSSM with RPV:} MSSM-like models can be realised with stacks of intersecting magnetised D7-branes wrapped around internal 4-cycles whose size determines the values of the SM gauge couplings \cite{Maharana:2012tu}. In these constructions $U(1)_{B-L}$ is a gauge symmetry at the fundamental level but it arises as an effective global symmetry below the St\"uckelberg mass of the corresponding gauge boson which is around the string scale. This effectively global $U(1)_{B-L}$ symmetry is exact at perturbative level but it gets broken by non-perturbative effects like stringy instantons. Some of these instantons break the continuous $U(1)_{B-L}$ symmetry down to a discrete $\mathbb{Z}_2$ symmetry which can be identified with $R$-parity \cite{Ibanez:2006da}, while other instanton contributions (with a different zero mode structure) would induce tiny $R$-parity violating couplings proportional to $e^{-S_{\rm inst}}\ll 1$, where $S_{\rm inst}$ is the instanton action \cite{Ibanez:2007rs}. 
\end{itemize}

\subsection{A string model for superheavy dark matter}

The generic features of 4D string compactifications described in Sec. \ref{GenFeat} might point towards an interesting scenario of superheavy decaying DM. In fact, Ref. \cite{Allahverdi:2020uax} has shown that high scale SUSY breaking and dilution from moduli domination can give rise to the right relic abundance of a superheavy WIMP $\chi$ with $m_\chi \sim 10^{10}$ GeV. 

The model is built within type IIB Large Volume Scenarios (LVS) where the complex structure moduli and the dilaton are stabilised by 3-form fluxes, while the K\"ahler moduli are fixed by combinations of perturbative (in both $\alpha'$ and $g_s$) and non-perturbative effects \cite{Balasubramanian:2005zx, Cicoli:2008va}. The minimal model involves 3 K\"ahler moduli (focusing on the canonically normalised modes): the overall volume modulus $\phi$ and two blow-up modes, $\sigma$ and $\rho$. One of them, which we will identify with $\sigma$, drives inflation as in K\"ahler moduli inflation \cite{Conlon:2005jm, Cicoli:2023njy} and it is wrapped by a hidden sector stack of D7-branes, while the other, $\rho$, is a spectator field during inflation and it is wrapped by the MSSM stack of D7-branes. During inflation, the volume modulus $\phi$ is shifted from its late-time minimum, which we set at $\phi=0$, by an amount of order $Y_\phi = \phi_0/M_p$ (where $M_p\simeq 10^{18}$ GeV is the Planck scale), and so drives an EMD epoch after the end of inflation \cite{Cicoli:2016olq}. 

The moduli break SUSY and generate a non-zero gravitino mass $m_{3/2} \simeq M_p / \mathcal{V} \ll M_p$, where $\mathcal{V}$ is the extra-dimensional volume in units of the string length. Their gravitational interaction with the MSSM induces soft SUSY breaking mass terms for squarks, sleptons and gauginos. At the Kaluza-Klein scale $M_{\rm KK} \simeq M_p / \mathcal{V}^{2/3}$, all SUSY scalars have a common mass given by $m_0$, while all gaugino masses are given by $M_{1/2}$. The mass spectrum at $M_{\rm KK}$ looks like:
\begin{equation}
m_\phi \sim m_{3/2} \,\sqrt{\frac{m_{3/2}}{M_p}} \ll 
m_0 = \frac{m_{3/2}}{\sqrt{2}} \simeq M_{1/2} = m_{3/2} \simeq m_\sigma \simeq m_\rho\,,
\label{MassScales}
\end{equation}
where to work out the exact coefficient of proportionality between the soft terms and $m_{3/2}$ we actually considered a slightly more involved situation where the MSSM is built with two stacks of intersecting and magnetised D7-branes wrapped around two intersecting blow-up cycles; one blow-up modulus is stabilised by setting a Fayet-Iliopoulos term to zero, while the other is fixed by an ED3-instanton.

The requirement of reproducing the right amount of density perturbations during inflation fixes the gravitino mass around $10^{10}$ GeV, corresponding to volumes of order $\mathcal{V}\sim 10^6$. This corresponds also to the typical mass scale of all SUSY particles and the two moduli $\sigma$ and $\rho$. On the other hand, the volume modulus is much lighter since $m_\phi \sim 10^7$ GeV. 

At the end of inflation, an initial reheating epoch is due to the decay of the inflaton $\sigma$ which decays mainly into hidden sector degrees of freedom since its coupling to MSSM fields is geometrically suppressed by the distance in the extra-dimensions between the inflaton 4-cycle and the 4-cycle supporting the MSSM D7-stack. Hence the production rate of MSSM particles, and in particular of the LSP, from the inflaton decay is very small since \cite{Allahverdi:2020uax}:
\begin{equation}
\frac{\Gamma_{\sigma \to {\rm MSSM}}}{\Gamma_{\sigma \to {\rm hid}}} \sim \frac{1}{\mathcal{V}^2}\ll 1\,.
\label{GammaRatio}
\end{equation}
In fact, the DM number density $n_\chi$ at inflaton decay reads (with $\Gamma_\sigma = \Gamma_{\sigma \to {\rm hid}}+ \Gamma_{\sigma \to {\rm MSSM}} \simeq \Gamma_{\sigma \to {\rm hid}})$:
\begin{equation}
n_\chi = n_\sigma \,\frac{\Gamma_{\sigma \to {\rm MSSM}}}{\Gamma_{\sigma }}\,{\rm Br}_{\rm odd}\,,   
\label{nchiin}
\end{equation}
where the inflaton number density is $n_\sigma \simeq 3 \Gamma_{\sigma \to {\rm hid}}^2 M_p^2/m_\sigma$, while ${\rm Br}_{\rm odd}$ is the branching ratio for the inflaton decay into $R$-parity odd particles which then decay into DM. 

The decay of $\rho$ can be safely ignored since this modulus would decay when it is not dominating the energy density. On the other hand, the volume mode $\phi$ has time to come to dominate the energy density (when $H=H_{\rm dom}\simeq Y_\phi^4\, \Gamma_\sigma$) before decaying. Its decay cannot reproduce any DM particles since $\phi$ is lighter than any superpartner, but it dilutes the $\chi$-particles produced at inflaton decay. To get the final DM relic abundance we need therefore to redshift (\ref{nchiin}) during the radiation-dominated (RD) epoch from $\sigma$-decay to the onset of $\phi$-domination, which amounts to a multiplicative factor $(H_{\rm dom}/\Gamma_{\sigma })^{3/2}$, and during the matter-dominated epoch from $\phi$-domination to $\phi$-decay, which yields another multiplicative factor of order $(\Gamma_\phi/H_{\rm dom})^2$. Putting all these factors together, we end up with \cite{Allahverdi:2020uax}:
\begin{equation}
\frac{n_\chi}{s} \simeq 10^{-3} \frac{1}{Y_\phi^2}\,\frac{\Gamma_{\sigma \to {\rm MSSM}}}{\Gamma_{\sigma }}\,\frac{T_{\rm rh}}{m_\sigma}\,,
\label{ThRelic}
\end{equation}
where we have normalised by the entropy density $s= 2\pi^2 g_* T_{\rm rh}^3/45$ at the reheating temperature from $\phi$-decay $T_{\rm rh}\sim \sqrt{\Gamma_\phi M_p}\sim m_\phi\sqrt{m_\phi/M_p}$. The relic density (\ref{ThRelic}) has to match the observed value \cite{Aghanim:2018eyx}:
\begin{equation}
\left(\frac{n_\chi}{s}\right)_{\rm obs} \simeq 4.2\times 10^{-10}\left(\frac{1\,{\rm GeV}}{m_\chi}\right). 
\label{ObsRelic}
\end{equation}
Reference \cite{Allahverdi:2020uax} has performed a detailed analysis, showing that (\ref{ThRelic}) and (\ref{ObsRelic}) indeed match for $m_\chi \sim 10^{10}$ GeV. This can be easily seen by noticing that, for such a large mass scale, $\left(n_\chi/s\right)_{\rm obs} \sim 10^{-20}$, which is reproduced by (\ref{ThRelic}) for $Y_\phi \sim 0.1$ (as computed in \cite{Cicoli:2016olq}), $\Gamma_{\sigma\to {\rm MSSM}}/\Gamma_\sigma \simeq 10^{-12}$ (from (\ref{GammaRatio}) for $\mathcal{V}\sim 10^6$) and $T_{\rm rh}/m_\sigma \sim \left(m_\phi/m_\sigma\right)\sqrt{m_\phi/M_p} \sim 10^{-7}$ (from (\ref{MassScales}) and $\mathcal{V}\sim 10^6$). 

\begin{figure}[t!]
    \centering
    \includegraphics[width=0.7\textwidth]{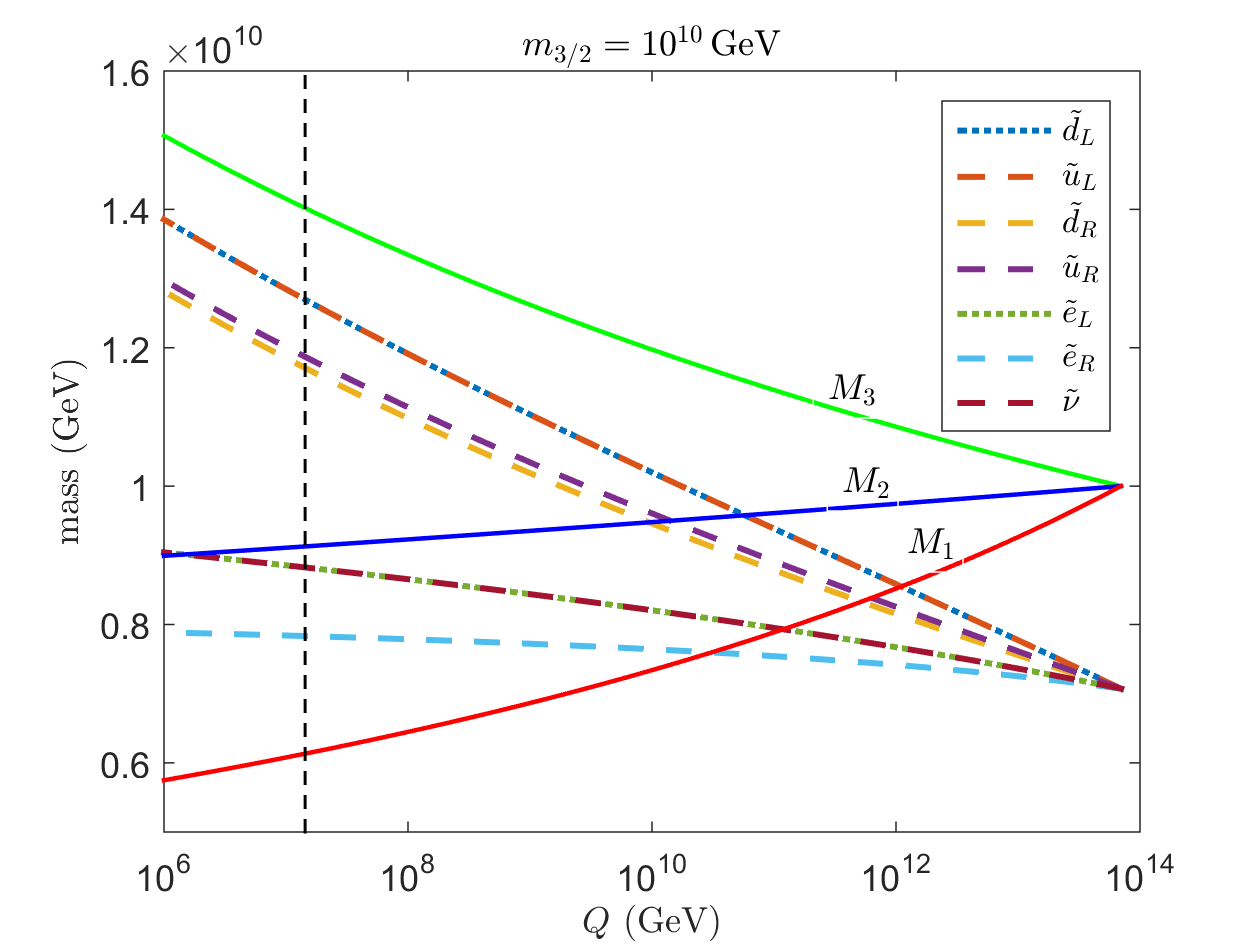}
    \caption{One-loop RGE running of the gaugino and scalar masses as a function of energy scale $Q$, with boundary conditions as described in the text. $M_1$, $M_2$, and $M_3$ depict the Bino, Wino and gluino, respectively, while the scalars appear as shown. The vertical dashed line marks the modulus mass $m_\phi \sim 10^7$~GeV. }
    \label{fig:RGE}
\end{figure}
Finally, we perform the RGE evolution of the gaugino masses and SM gauge couplings at one-loop order to estimate the mass spectrum of SUSY particles at lower energies in the model described above. 
We begin the evolution at the Kaluza-Klein scale with the following boundary conditions: $M_{1/2} = m_{3/2}$, $m_0 = m_{3/2}/\sqrt{2}$, with the gravitino mass set to $m_{3/2} = 10^{10}$~GeV and the gauge couplings given by their approximate unification value $g_a = 0.7$. From there we run down to the mass scale of the volume modulus $m_\phi \sim 10^7$~GeV to obtain the gaugino and scalar masses at low energies (see App. \ref{app:RGE} for details). 
In Fig.~\ref{fig:RGE} we depict the resultant RGE running with the scalars labeled in the legend and the Bino, Wino, and gluino shown by $M_1$, $M_2$, and $M_3$ respectively. The squarks receive strong contributions from the gluino mass, pushing their masses higher than their initial value, while the slepton masses run more mildly, remaining closer to their initial value. We see that in this example the gluino is the heaviest sparticle, while the Bino is the LSP.

\subsection{Decaying dark matter and three-body decay signatures}
\label{sec:spectrum}

As explained in Sec. \ref{GenFeat}, an important generic feature of MSSM model-building in string theory is the presence of exponentially small RPV couplings generated by stringy instantons \cite{Ibanez:2006da,Ibanez:2007rs}. These couplings can therefore make the LSP unstable, leading to a decaying DM scenario if the lifetime of the $\chi$-particles is not smaller than the age of the Universe.  

\begin{figure}[t!]
    \centering
    \includegraphics{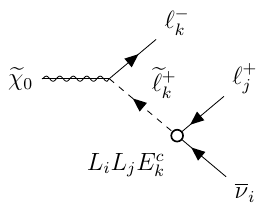}
    \includegraphics{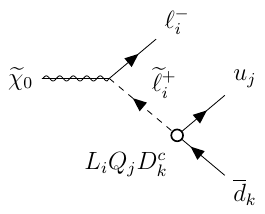}
    \includegraphics{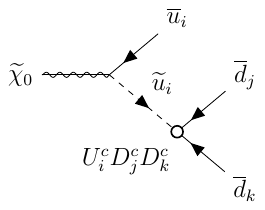}
    \caption{Feynman diagrams illustrating the decay of the LSP $\widetilde{\chi}_0$ into three SM fermions, mediated by a sfermion $\widetilde{f}$.
    The RPV vertex is highlighted by the white dots.}
    \label{fig:neutralino_rpv}
\end{figure}

To take into account RPV, we shall include the following trilinear contributions to the superpotential~\cite{Baltz:1997gd}:\footnote{Note that there are other possibilities to introduce RPV, see \cite{Barbier:2004ez}, but these are not considered here.} 
\begin{equation}
\label{eq:w_rpv}
    W_{\rm RPV} = \lambda LLE^c + \lambda' LQD^c + \lambda'' U^c D^c D^c \,,
\end{equation}
where the first two terms violate lepton number, while the third one violates baryon number.
In more detail, \eqref{eq:w_rpv} reads:
\begin{eqnarray}
\label{eq:explicit}
    W_{LLE} & = & \epsilon^{\sigma\rho} \left( \lambda_{ijk} L_{i\sigma} L_{j\rho} E^c_{k} \right)\, ,\\
   W_{LQD} & = & \epsilon^{\sigma \rho} \left(  \lambda'_{ijk} L_{i\sigma} Q_{j\rho \alpha} D^c_{k\alpha}\right)\,, \\
  W_{UDD} & = & 2 \epsilon^{\alpha \beta \gamma} \lambda''_{ijk} U^c_{i\alpha} D^c_{j\beta} D^c_{k \gamma} \, ,
\end{eqnarray}
where $i,j,k$ are generation indices ($k>j$), $\sigma,\rho$ are $SU(2)_L$ indices, while $\alpha,\beta,\gamma$ are $SU(3)$ colour triplet indices. These terms give rise to an interesting phenomenology, see {\it e.g.} Refs.~\cite{Dawson:1985vr,Barbier:2004ez} for details. In particular, these vertices make a sfermion decay into two fermions, allowing for instance the sneutrino or the neutralino, $\widetilde{\chi}_0$, to decay. Indeed in the MSSM, the neutralino-sfermion-fermion vertex is given by:
\begin{equation}
\label{eq:vertex}
    \mathcal{L}_{\chi f \widetilde{f}} = \bar{\chi}_0 \left( g_{\chi f_{ik}}^L P_L + g_{\chi f_{ik}}^R P_R
    \right) f_{i\alpha} \widetilde{f}^\ast_{k \alpha} + h.c. \,,
\end{equation}
where $k$ specifies the sfermion mass eigenstate, while $P_L$ and $P_R$ stand for left and right projectors. The couplings are proportional to the sfermion mixing matrix and to all the neutralino eigenstates (Bino, Wino and Higgsino). The sfermion in Eq.~\eqref{eq:vertex} is then allowed to decay via one of the RPV terms, in Eq.~\eqref{eq:explicit}, and gives as a result a three-body final state, as illustrated in Fig.~\ref{fig:neutralino_rpv}. The $LLE$ term produces a decay into three leptons, one of which should be a neutrino, while the other two terms produce predominantly quarks. 

In what follows, we will consider the SUSY neutralino, $\widetilde{\chi}_0$, as the DM candidate, while in App.~\ref{app:snudecay} we will discuss the case of a sneutrino DM candidate. The decay of the neutralino will take place regardless of its composition. Note, however, that for the Higgsino case, since the coupling is proportional to the mass of the fermion, there will be a strong suppression due to the Yukawa coupling. From now on we will work in the pure Bino approximation (see Fig.~\ref{fig:RGE}) as discussed below.
The $\widetilde{\chi}_0$ decay due to the RPV terms is given by \cite{Dawson:1985vr,Baltz:1997ar,Barbier:2004ez}: 
\begin{eqnarray}
\label{eq:gamma1_neutralino}
\Gamma_{\widetilde{\chi}_0}^{LLE} & = & \lambda^2_{ijk}\, \frac{\alpha(m^2_{\widetilde{\chi}_0})}{128 \pi^2} \frac{m^5_{\widetilde{\chi}_0}}{m^4_{\widetilde{f}}}\,,\\
\label{eq:gamma2_neutralino}
\Gamma_{\widetilde{\chi}_0}^{LQD} & = & \lambda^{'2}_{ijk}\, \frac{3 \alpha(m^2_{\widetilde{\chi}_0})}{128 \pi^2}  \frac{m^5_{\widetilde{\chi}_0}}{m^4_{\widetilde{f}}}\,,\\
\label{eq:gamma3_neutralino}
\Gamma_{\widetilde{\chi}_0}^{UDD} & = & \lambda^{'' 2}_{ijk}\, \frac{3\alpha(m^2_{\widetilde{\chi}_0})}{64 \pi^2} \frac{m^5_{\widetilde{\chi}_0}}{m^4_{\widetilde{f}}}\,,
\end{eqnarray}
where $\alpha$ is the fine-structure constant at the scale of the DM mass. All decay widths have been computed in the limit of degenerate heavy sfermion masses. The $LQD$ and $UDD$ vertices that depend on coloured particles scale as $N_c$ and $N_c!$ respectively, with $N_c=3$ being the colour factor.

The neutralino lifetime $\tau_{\widetilde{\chi}_0}= \Gamma_{\widetilde{\chi}_0}^{-1}$ should be larger than $\tau_{\widetilde{\chi}_0} \gtrsim 10^{26}$ s to be compatible with current bounds on decaying DM, see \cite{Chianese:2021htv, Chianese:2021jke, Esmaili:2015xpa, Arguelles:2022nbl, Murase:2012xs, Esmaili:2012us, Cohen:2016uyg, Kalashev:2016cre, HAWC:2017udy, Kachelriess:2018rty, Ishiwata:2019aet, Kalashev:2020hqc, Guepin:2021ljb, LHAASO:2022yxw, IceCube:2023ies, Fiorillo:2023clw}, and to be larger than the age of the Universe. For $m_{\widetilde{\chi}_0}\simeq m_{\widetilde{f}}\simeq 10^{10}$ GeV, this would require very small RPV couplings of order $\lambda_{ijk} \lesssim 10^{-26}$, in the case of the $LLE$ RPV term. Interestingly, these values are in the right ballpark expected in 4D string models. In fact, RPV couplings are generated by ED3-instantons wrapped around the 4-cycle $\tau_{\rm SM}$ supporting the MSSM D7-stack, and so $\lambda_{ijk} \sim e^{-S_{\rm inst}}$ where the instanton action is given by $S_{\rm inst} = 2\pi \tau_{\rm SM}$. In LVS models $\tau_{\rm SM}$ is a blow-up modulus which is typically stabilised at values of order the inverse of the string coupling $g_s$, $\tau_{\rm SM}\simeq g_s^{-1}$. Therefore we expect $\lambda_{ijk} \sim e^{-2\pi/g_s}$ which, for values of the string coupling $g_s\lesssim 0.1$ that keep the effective field theory in the perturbative regime (which can be guaranteed by an appropriate choice of 3-form flux quanta), would give indeed $\lambda_{ijk} \sim e^{-20\pi} \lesssim 5\times 10^{-28}$. 

\section{Astrophysical signatures of superheavy decaying dark matter}
\label{sec:expconstr}

Superheavy DM particles $\chi$ distributed in galactic and extra-galactic structures can decay into ordinary matter by producing Ultra-High-Energy (UHE) particles. Here we focus on UHE neutrino and gamma-ray fluxes produced by DM decays. UHE charged cosmic rays are also an interesting probe for DM decay. However, their understanding is subject to larger astrophysical uncertainties, because of their propagation in the galaxy, with respect to neutrinos and gamma rays which trace the source. Our analysis on neutrino and gamma-ray data is based on Refs.~\cite{Chianese:2021htv,Chianese:2021jke} which we refer the reader to for further details.

\subsection{Theoretical astrophysical fluxes}
The neutrino and gamma-ray fluxes produced by DM decay originate in principle from both galactic and extra-galactic sources. However, the gamma-ray flux coming from extra-galactic production is expected to be substantially attenuated, due to the absorption by $\gamma\gamma$ scattering happening at high energies and high redshift, and can be subsequently neglected. Additionally, the secondary photon emission from Inverse Compton scatterings of electrons and positrons contributes to gamma-ray observations at energies $E_\gamma \lesssim 10^5~{\rm TeV}$, and therefore can be safely neglected in our analysis based on UHE gamma rays. Hence, the dominant component at $E_\gamma \gtrsim 10^5~{\rm TeV}$ is the prompt galactic emission which gives the following expected flux:
\begin{equation}
    \frac{{\rm d}\Phi_{\gamma}}{{\rm d}E_\gamma {\rm d}\Omega} = \frac{1}{4\pi m_\chi \tau_\chi} \frac{\mathrm{d}N_\gamma}{\mathrm{d}E_\gamma} \int_0^{+\infty} \mathrm{d}s \, \rho_\chi[r(s, \ell, b)] \exp[-\tau_{\gamma\gamma}(E_\gamma,s,b,\ell)]\,,
    \label{eq:gamma_flux_galactic}
\end{equation}
where $\tau_{\gamma\gamma}$ represents the total optical depth of photons to $\gamma\gamma$ collisions and $\rho_\chi = \rho_s [r/r_r(1+r/r_s)^2]^{-1}$ with $r_s = 24~{\rm kpc}$ and $\rho_s=0.18~{\rm GeV\, cm^{-3}}$ is the Navarro-Frenk-White galactic DM density profile~\cite{Navarro:1995iw, Cirelli:2010xx}. The optical depth takes into account the photon pair production from the interaction of the gamma rays with CMB radiation, starlight and infrared light (see Refs.~\cite{Esmaili:2015xpa,Chianese:2021jke} for details on its computation). While the CMB photon density is homogeneous and thermally defined by the CMB temperature ($\simeq 2.348\times10^4~{\rm eV}$), the photon density of starlight and infrared light strongly depends on the position in the Milky Way. For the latter, we take the photon density from the {\ttfamily GALPROP} code~\cite{galprop}.
The density profile is expressed as a function of the distance $r$ from the center of our galaxy. The integration is performed along the line of sight $s$ from the Earth, so that $r$ can be written as a function of the colongitude $\ell$ and colatitude $b$, according to the following expression:
\begin{equation}
    r = \sqrt{s^2 + r_\odot^2 - 2 s r_\odot \cos\ell \cos b}\,,
    \label{eq:galactocentric_radial_coordinate}
\end{equation}
where we identify $r_\odot = 8.5~\mathrm{kpc}$ as the distance between the Earth and the galactic center.

Current UHE gamma-ray observatories place bounds on the diffuse gamma-ray flux, which can be linked to the integral gamma-ray flux, averaged over the field of view of each experiment $\Omega_{\rm exp}$ (see supplementary Tab.~I in Ref.~\cite{Arguelles:2022nbl}), which takes the following expression:
\begin{equation}
    \Phi^\chi_\gamma(E_\gamma) = \frac{1}{\Omega_{\rm exp}} \int_{E_\gamma}^\infty \mathrm{d}E'_\gamma \int_{\Omega_{\rm exp}} \mathrm{d}\Omega \frac{\mathrm{d}\Phi_\gamma}{\mathrm{d}E'_\gamma\mathrm{d}\Omega}\,.
    \label{eq:integral_gamma}
\end{equation}

Contrarily to gamma rays, neutrinos travel unimpeded and the neutrino flux from DM decays has both a galactic and extra-galactic component. The first one is given by:
\begin{equation}
    \frac{{\rm d}\Phi^\textup{gal}_{\nu_\alpha + \overline{\nu}_\alpha}}{{\rm d}E_\nu {\rm d}\Omega} = \frac{1}{4\pi m_\chi \tau_\chi} \frac{\mathrm{d}N_\alpha}{\mathrm{d}E_\nu} \int_0^{+\infty} \mathrm{d}s \, \rho_\chi[r(s, \ell, b)]\,,
    \label{eq:neutrino_flux_galactic}
\end{equation}
where $\alpha$ is the flavour of the (anti)neutrino considered.
The important feature of extra-galactic neutrinos is that their absorption in the inter-galactic medium is negligible. Hence they produce a sizeable flux that can be written as:
\begin{equation}
    \frac{{\rm d}\Phi^\textup{ex-gal}_{\nu_\alpha + \overline{\nu}_\alpha}}{{\rm d}E_\nu {\rm d}\Omega} = \frac{\Omega_\chi\, \rho_\textup{c}}{4\pi m_\chi \tau_\chi} \int_0^{+\infty} \frac{\mathrm{d}z}{H(z)} \frac{\mathrm{d}N_\alpha}{\mathrm{d}E_\nu^\prime}\biggr\rvert_{E_\nu^\prime = E_\nu(1+z)},
    \label{eq:neutrino_flux_extra-galactic}
\end{equation}
where $H$ is the Hubble parameter, $\rho_\textup{c}$ is the critical density and $\Omega_\chi$ is the DM relic density, according to the latest Planck data~\cite{Planck:2018vyg}. This flux depends on the cosmological redshift $z$.

Both galactic and extra-galactic neutrinos travel a long distance before reaching Earth and they oscillate very rapidly with respect to the propagation scale, so that the detected neutrinos have a composition of flavour eigenstates given by averaged fast oscillations.
This feature is taken into account by defining a total flux given by the sum of each flavour flux.
Moreover, experiments are not sensitive to the neutrino arrival direction, but only to its average. We keep that into account by performing an integration over the solid angle.
As a result, the total combined flux is:
\begin{equation}
    \frac{\mathrm{d}\Phi^\chi_{3\nu}}{\mathrm{d}E_\nu} = \sum_\alpha \int_{4\pi} \mathrm{d}\Omega \biggl[\frac{{\rm d}\Phi^\textup{gal}_{\nu_\alpha + \overline{\nu}_\alpha}}{{\rm d}E_\nu {\rm d}\Omega} + \frac{{\rm d}\Phi^\textup{ex-gal}_{\nu_\alpha + \overline{\nu}_\alpha}}{{\rm d}E_\nu{\rm d}\Omega} \biggr]\,.
    \label{eq:neutrino_flux_total}
\end{equation}
Considering the angle-averaged flux means to neglect neutrino flux anisotropies which are particularly relevant for galactic neutrinos in the direction of the galactic center.
This implies that the combined flux expressed by Eq.~\eqref{eq:neutrino_flux_total} allows us to understand only the flux normalization, staying conservative on the angular distribution.

The neutrino and gamma-ray fluxes in Eqs.~\eqref{eq:gamma_flux_galactic},~\eqref{eq:neutrino_flux_galactic} and ~\eqref{eq:neutrino_flux_extra-galactic} depend on the energy spectrum, ${\mathrm{d}N_{\nu,\gamma}}/{\mathrm{d}E_{\nu,\gamma}}$, that denotes the energy distribution of neutrinos and photons deriving from the three-body or two-body DM decays. This is a central key feature of this paper that needs to be carefully evaluated to have proper estimates of the sensitivity of current and future experiments. We will discuss how we properly compute this quantity in Sec.~\ref{sec:spectra}, after having introduced the experimental neutrino and gamma-ray probes, as well as the corresponding statistical approach we adopt to compute the limits on the DM lifetime $\tau_\chi$ for different DM masses $m_\chi$.

\subsection{Future neutrino telescopes}

When neutrinos reach the Earth, they can be detected by the hadronic or electromagnetic showers that they create when interacting in a medium.
Many new experiments will be available in the near future to collect these signals for UHE neutrinos.
In general, the showers originating from UHE neutrinos are characterized by emissions in the radio spectrum, a phenomenon known as the Askaryan effect.
However, since the neutrino fluxes at these high energies are low, the experiments should be equipped with a large interaction volume, so that the main mediums that have been considered so far are air, water and ice.
In the following we give a brief description of the main experiments at play in the next decades:
\begin{description}
    \item[RNO-G~{\normalfont\cite{RNO-G:2020rmc}}] the Radio Neutrino Observatory in Greenland is a neutrino telescope exploiting an ice medium for neutrino interaction.
    UHE neutrinos can produce hadronic or electromagnetic showers inside the ice which can then be collected by the antennas equipped to each one of the 35 stations the experiment is made of.
    \item[IceCube-Gen2~{\normalfont\cite{IceCube:2019pna,IceCube-Gen2:2020qha}}] the next generation of the IceCube experiment is located in Antarctica, exploiting the natural ice for neutrino interaction.
    The detection of the showers relies on the radio array made of 200 stations.
    \item[GRAND~{\normalfont\cite{GRAND:2018iaj}}] the Giant Radio Array for Neutrino Detection is a planned observatory sensitive to cosmic rays, gamma rays and neutrinos.
    Its main strategy for neutrino detection focuses on Earth-skimming underground $\tau$ neutrinos: they are astrophysical $\nu_\tau$ crossing the Earth medium at a very small angle with the surface, or passing through mountains.
    In this environment, neutrinos can interact with the medium by charged-current interaction and can produce $\tau$ leptons which can eventually decay outside of the rock, producing an electromagnetic shower which can be detected by the large number of antennas foreseen in the design of GRAND. Here, we consider two different GRAND configurations with 10k and 200k radio antennas, respectively.
\end{description}

Superheavy decaying DM offers a natural source of neutrinos that can propagate through the galaxy and reach the Earth.
In particular, given a DM candidate of mass $m_\chi$ and lifetime $\tau_\chi$, we can compute the expected number of neutrino events in a given telescope, assuming an observation period of $T_\textup{obs} = 3~{\rm yr}$:
\begin{equation}
    N^\chi_\nu = T_\textup{obs} \int_0^{E_{\rm max}} \mathrm{d}E_\nu \frac{\mathrm{d}\Phi_{3\nu}^\chi}{\mathrm{d}E_\nu} A_\textup{eff}(E_\nu)\,,
    \label{eq:Nnu}
\end{equation}
where $E_{\rm max}=m_\chi/2$ is the maximum neutrino energy achieved in the DM decay. The neutrino flux from DM decay has been previously defined in Eq.~\eqref{eq:neutrino_flux_total}, while we define the effective area of the experiment with $A_\textup{eff}$.
This can be defined through the flux sensitivity $S(E_\nu)$, which is the quantity provided by the experimental collaborations, and the observation time:
\begin{equation}
    A_\textup{eff}(E_\nu) = \frac{2.44\, E_\nu}{4\pi \log (10) S(E_\nu) T_\textup{obs}}\,,
\end{equation}
where $2.44$ is the number of neutrino events for each energy decade. 

Following Ref.~\cite{Chianese:2021htv}, we will consider two main contributions (taken from Ref.~\cite{GRAND:2018iaj}) for the UHE neutrino flux coming from astrophysical sources:
\begin{itemize}
    \item {\bf Cosmogenic neutrinos:} They originate from collisions between high-energy cosmic rays and the CMB. Their contribution to the flux of UHE neutrinos is affected by important theoretical uncertainties and, to be conservative, we will consider the most optimistic estimate on this contribution in order to obtain the weakest limits on DM;
    \item {\bf Pulsar neutrinos:} The extreme astrophysical environment of newborn pulsars might give rise to one of the highest UHE neutrino fluxes.
\end{itemize}

Such astrophysical models predict the observation of a certain number of astrophysical neutrino events, $N^{\rm astro}_\nu$, which can be computed from the neutrino flux through an equation analogous to Eq.~\eqref{eq:Nnu}. Hence, we forecast conservative limits on the DM lifetime by assuming the observation of $N^{\rm obs}_\nu$ neutrino events (with $N_{\rm obs}$ being a Poissonian stochastic variable with mean $N^{\rm astro}_\nu$) and by taking the test statistic:
\begin{equation}
    \mathrm{TS}(m_\text{DM},\tau_\text{DM})=
    \begin{cases}~0 & \text{for} \quad N^\chi_\nu < N^\mathrm{obs}_\nu \\
    -2\ln\left(\frac{\mathcal{L}(N^\mathrm{obs}_\nu | N^\chi_\nu)}{\mathcal{L}(N^\mathrm{obs}_\nu|N^\mathrm{obs}_\nu)}\right)& \text{for} \quad  N^\chi_\nu\geq N^\mathrm{obs}_\nu \end{cases} \,,
\end{equation}
where the likelihood $\mathcal{L}$ is assumed to be a Poisson distribution. According to this test statistic~\cite{Cowan:2010js}, we can place future bounds on the DM lifetime $\tau_\chi$ at 95\% CL by taking ${\rm TS}(m_\chi, \tau_\chi) = 2.71$.

\subsection{Current gamma-ray experiments}

UHE photons and cosmic rays reaching the Earth's atmosphere will interact producing extensive air showers (EAS), generating cascades of charged particles that can leave traces in the detectors placed on the Earth's surface.
Gamma-ray experiments usually employ one or more scintillation arrays or huge \v{C}herenkov stations.
The latter aim at detecting the \v{C}herenkov light coming from the particles in the EAS travelling through the atmosphere, that is collected by photomultiplier arrays.
One of the main obstacles for the measurement of gamma rays with ground-based telescopes is the high level of residual charged cosmic ray background.
EAS can not be distinguished by background charged cosmic rays based on directional information.

In the present study, we focus on the following current and decommissioned experiments which have placed several bounds on the UHE gamma-ray diffuse flux above $10^5~{\rm GeV}$ (see Fig.~1 in Ref.~\cite{Chianese:2021jke}).
\begin{description}
    \item[CASA-MIA~{\normalfont\cite{CASA-MIA:1997tns}}] the Chicago Air Shower Array-MIchigan Muon Array experiment was a surface array of scintillation counters located in Utah (USA), able to infer the properties of EAS and it was active until 1998. The main detection strategy relied on discriminating gamma-ray primary showers from hadronic (cosmic ray)-generated ones, by identifying muon-less or muon-poor events. This principle is based on the fact that muons originate mainly from charged pion and kaon decays which are abundant in a hadronic shower, while the muon content in an electromagnetic shower is mainly coming from photoproduction events whose interaction rate is suppressed with respect to nucleus-nucleus events.
    \item[KASCADE and KASCADE-GRANDE~{\normalfont\cite{KASCADEGrande:2017vwf}}] the KArlsruhe Shower Core and Array DEtector was located at the Karlsruhe Institute of Technology (Germany) and operated until the end of 2012. It was made of three detector systems: a large field array, a muon tracking detector and a central detector. The experiment's aim was to measure the electromagnetic, muonic and hadronic components of EAS.
    \item[TA~{\normalfont\cite{TelescopeArray:2018rbt}}] the Telescope Array experiment operates in Utah (USA) since 2008 and it is made of a surface detector with plastic scintillators and three fluorescence detectors. It obtained limits on diffuse photon flux with energies greater than $\mathrm{EeV}$.
    \item[PAO~{\normalfont\cite{PierreAuger:2016kuz}}] the Pierre Auger Observatory is a surface detector array of \v{C}herenkov stations, sensitive to the electromagnetic, muonic and hadronic components. An additional fluorence detector completes the setup to observe the longitudinal development of the shower. It operates in Argentina and it performed a search for UHE photons with energies above $\mathrm{EeV}$.
\end{description}

Given the upper bounds on the angle-averaged integral gamma-ray flux from each experiment, we compute the limits on DM by simply performing a $\chi^2$ analysis using the test statistic:
\begin{equation}
    \mathrm{TS}(m_\chi,\,\tau_\chi) = \sum_i\left[\frac{\Phi^\chi_{\gamma}(E_{\gamma,i})-\Phi_{\gamma,\,i}^\mathrm{data}}{\sigma_i}\right]^2,
    \label{eq:xi2}
\end{equation}
where the index $i$ runs over the experimental energy bins, $\Phi_\gamma^{\rm data}=0$ (no observation of photons) and $\sigma_i$ is the corresponding standard deviation at 68\% CL of each data point. The limits on the DM lifetime at 95\% CL can be obtained by taking ${\rm TS}=2.76$, a threshold value which derives from the constraint that the gamma-ray measurements cannot be negative~\cite{Chianese:2021jke}.

\section{Methodology and energy spectra}
\label{sec:spectra}

The DM decay rates and final states have been computed using the UFO model that is publicly available in the \fr~\cite{Alloul:2013bka,Degrande:2011ua} database.\footnote{The RPV model can be found at this webpage \url{https://feynrules.irmp.ucl.ac.be/wiki/RPVMSSM}} 
It encodes the MSSM with trilinear vertices for RPV interactions and it is based on~\cite{Fuks:2012im,Barbier:2004ez}. It allows for the choice of the SUSY mass spectrum, in such a way that either the neutralino or sneutrino can be set as the LSP. 

\begin{table}[tp]
\caption{Three-body neutralino final states induced by the RPV vertices in Eqs.~\eqref{eq:explicit}.
The indices $i,j,k$ label the fermion generations.}
\centering
\begin{tabular}{cccc}
\toprule
 & \multicolumn{3}{c}{Couplings / channels} \\
 &~~$\lambda_{ijk}$ / LLE channel~~& $~~\lambda_{ijk}^{\prime}$ / LQD channel~~&~~$\lambda_{ijk}^{\prime\prime}$ / UDD channel~~\\
\midrule
\multirow{2}{*}{$\widetilde{\chi}_0$} & $\ell_i^+ \overline{\nu}_j \ell_k^-$, $\ell_i^- \nu_j \ell_k^+$, & $\ell_i^+ \overline{u}_j d_k$, $\ell_i^- u_j \overline{d}_k$, & \multirow{2}{*}{$\overline{u}_i \overline{d}_j \overline{d}_k$, $u_i d_j d_k$} \\
 & $\overline{\nu}_i \ell_j^+ \ell_k^-$, $\overline{\nu}_i \ell_j^- \ell_k^+$ & $\overline{\nu}_i \overline{d}_j d_k$, $\nu_i d_j \overline{d}_k$ & \\
\bottomrule
\end{tabular}
\label{tab:RPVdecays}
\end{table}

The SUSY mass spectrum has been chosen in such a way that the gauginos and sfermions are all at the same mass scale, which is given by the gravitino mass, as explained in Sec.~\ref{sec:model}. In order to optimise the generation, we have discarded irrelevant and time-consuming contributions in the decay matrix element computation. In particular, we have set to a much heavier mass scale all SUSY particles that do not participate in the decay.

The UFO model files have been loaded in \aNLO~\cite{Alwall:2014bza} to obtain the decay rates, including the three-body decays of the neutralino. The kinematics of the final states is fixed for a two-body decay, while it is not for a three-body final state. In this case the distribution in energy of the final state particles of the hard process needs to be simulated accurately. Tab.~\ref{tab:RPVdecays} shows all possible final states for the neutralino decay. We have turned on one RPV vertex at a time, in order to study separately the three channels $LLE$, $LQD$ and $UDD$. 

Considering the case of neutralino DM, the $LLE$ vertex gives rise to three leptons of different flavour in the final state, one of them always being a neutrino, such as $\widetilde{\chi}_0 \to \mu^{+} e^{-} \nu_{\tau},\,  \widetilde{\chi}_0 \to e^+ e^- \nu_\mu$ and so on.  There are 36 different final states. We run \aNLO\ to simulate the hard process, asking for $10^6$ events per decay, which ensures a statistic large enough to sample accurately each single final state. We proceed in a similar manner for the $LQD$ and $UDD$ vertices, requiring $10^6$ events per decay. The former gives rise to final states with one lepton and two quarks, such as $\widetilde{\chi}_0 \to e^-\, u\, \overline{d},\, \widetilde{\chi}_0 \to \overline{\nu}_e\, s\, \overline{d}$, for a total of 108 final states. The third RPV vertex gives rise to 18 final states with different combinations of quarks, \eg~$\widetilde{\chi}_0 \to u\, s\, b,\, \widetilde{\chi}_0 \to \overline{b}\, \overline{t}\, \overline{s}, \, ...$. 

The hard process final states of the three-body neutralino decay are highly energetic, because we are considering DM particles with masses $m_\chi \simeq 10^{10}$ GeV.
Every final state will generate a shower, producing a cascade of new particles.
Given the high energy of the process, it is important to properly include electroweak corrections that affect the final energy spectra of neutrinos and gamma rays (for discussion, see~\cite{Ciafaloni:2010ti,Bauer:2020jay}).
Notice that the electroweak evolution is collinear, and therefore local. Hence, given the mother particle and its energy, it is enough to perform the evolution. This can be done consistently using the electroweak corrections obtained with the tool \hdmspectra~\cite{Bauer:2020jay}.

\hdmspectra\ provides electroweak-corrected spectra for $\gamma$, $\nu$'s and protons originating from the two-body annihilation (or decay) of a generic DM particle, for energies up to $10^{19}$ GeV:\footnote{In the case of decay, \hdmspectra\ simulates the process $\chi \to f \bar{f}$; using annihilation or decay is equivalent in the procedure we followed, and in our case we preferred to use annihilation because energy is sampled over $x=E/m_\chi$ instead of $x=2E/m_\chi$ used in decay, which would introduce several factors of $2$ in the following steps.}
\begin{equation}
    \chi \chi \to f \bar{f} \to~\mathrm{shower}\,,
\end{equation}
for which the hard spectrum is highly peaked.
However, in our case we need to compute the spectra for a three-body decay process, distributed accordingly to a continuum of energies, hence a convolution operation between the hard spectra and \hdmspectra\ results is necessary. We proceed as follows. For each RPV vertex, we sum over all the neutralino three-body decay processes, getting $P$ different particles overall, {\it e.g.}:
\begin{equation}
    \begin{rcases}
        \widetilde{\chi}_0 &\to e^{+} e^{-} \nu_{\mu}\\
        \widetilde{\chi}_0 &\to \mu^{+} e^{-} \nu_{\tau}\\
        \widetilde{\chi}_0 &\to \dots
    \end{rcases}
    \quad\Rightarrow \quad p_1 \dots p_P\,.
    \label{eq:example_spectra_decay_3body}
\end{equation}
The convolution operation requires us to consider each $p_i$, compute the electroweak shower spectra with \hdmspectra\ and sum over every particle. To avoid doing that for each event, we bin the particles in the hard spectra and do a convolution.
This is possible assuming that the particles belonging to each bin in the hard spectra have the same energy, and in particular equal to the mid value of the bin. This is true if the binning scheme is fine enough to have very small bins, which is realised by taking 1000 bins in our case.
The convolution is made by considering that each binned particle is coming from a \emph{fake} two-body annihilation with a proper center-of-mass energy to match the energy of the bin.
The result will be the electroweak-corrected spectra coming from the three-body decay of the neutralino. 

More precisely, the hard process spectrum generated by \aNLO~is stored in an event file in the LHE format \cite{Alwall:2006yp}, containing the different generated processes and kinematics for each event.
This file can be parsed to extract the hard spectra of the final states.
Each hard spectrum of the particle $p_i$, with $i \in [1, P]$, is a one-dimensional histogram with $M$ bins in the variable $x = E/m_{\widetilde{\chi}_0}$, where $E$ is the energy of the particle $p_i$.
Each bin $j$, with $j \in [1, M]$, of the hard spectrum of the particle $p_{i}$ will contain $n_{ij}$ particles and we have:
\begin{equation}
    \sum_{i=1}^{P} \sum_{j=1}^{M} n_{ij} = D\,,
\end{equation}
where $D$ is the total number of generated events for the decays in \eqref{eq:example_spectra_decay_3body}.
Furthermore, the histogram is normalised according to $D$, so that:
\begin{equation}
    \frac{n_{ij}}{D} = w_{ij}\qquad\text{and}\qquad
    \sum_{i=1}^{P} \sum_{j=1}^{M} w_{ij} = 1\,.
\end{equation}
$w_{ij}$ is the probability that the three-body decay of $\chi$ will yield the particle $p_i$ with energy fraction $x_j$ (belonging to the $j$-th bin).

Then we need to compute the shower spectrum for each of the three-body decays.
Every particle $p_i$ coming from the decay will produce a shower, generating several different particles:
\begin{equation}
    p_i \to s_1 \dots s_H\,,
    \label{eq:shower_pi_to_sk}
\end{equation}
where $H$ is the number of different particles types generated ($\gamma$, $\nu$'s, \dots), each one being a state $s_{k}$, with $k \in [1,H]$.
We remark that the process written in \eqref{eq:shower_pi_to_sk} happens for each particle in the histogram of $p_i$, where every bin $j$ is associated to the weight $w_{ij}$ and contains particles $(p_i)_j$ with energy $E_j = x_j m_{\widetilde{\chi}_0}$.
For each bin $j$, we run \hdmspectra\  simulating the following process:
\begin{equation}
    \chi_j \chi_j \to (p_i)_j (\bar{p}_i)_j \to s_1 \dots s_H\,,
\end{equation}
where $\chi_j$ is a \emph{fake} DM with a mass consistent with the energy of the particles $(p_i)_j$:
\begin{equation}
    m_{\chi_j} = x_j m_{\widetilde{\chi}_0} = E_j.
\end{equation}
Because \hdmspectra\ samples its output on the basis of $m_\chi$, with an array binned in $y_j = E/m_{\chi_j}$, we substitute: 
\begin{equation}
    y_j = \frac{E}{m_{\chi_j}} = x \frac{m_{\widetilde{\chi}_0}}{E_j}\,,
\end{equation}
to have results in terms $x$. 
\hdmspectra\ returns the values of $( {\mathrm{d}N_k}/{\mathrm{d}y_j} )_i$, which represents the electroweak-corrected spectra of the final state $s_k$, as coming from the shower of the hard spectrum particle $p_i$ with energy $E_j$.
This spectrum should be rescaled appropriately:
\begin{equation}
    \biggl( \frac{\mathrm{d}N_k}{\mathrm{d}x} \biggr)_{ij} = \frac{\mathrm{d}y_j}{\mathrm{d}x} \biggl( \frac{\mathrm{d}N_k}{\mathrm{d}y_j} \biggr)_i = \frac{m_{\widetilde{\chi}_0}}{E_j} \biggl( \frac{\mathrm{d}N_k}{\mathrm{d}y_j} \biggr)_i\,,
\end{equation}
 for every bin $j$ of the particle $p_i$ hard spectrum and then for any particle $p_i$ of the decay processes.
The total spectrum of the particle $s_k$ is eventually given by summing all of the contributions for each bin $j$ and for each particle $p_i$, weighting each result on the basis of the probability $w_{ij}$ discussed before:
\begin{equation}
    \frac{\mathrm{d}N_k}{\mathrm{d}x} = \sum_{i=1}^{P} \sum_{j=1}^{M} w_{ij} \biggl( \frac{\mathrm{d}N_k}{\mathrm{d}x} \biggr)_{ij}.
\end{equation}
\begin{figure}[t!]
    \centering
    \includegraphics[scale=1]{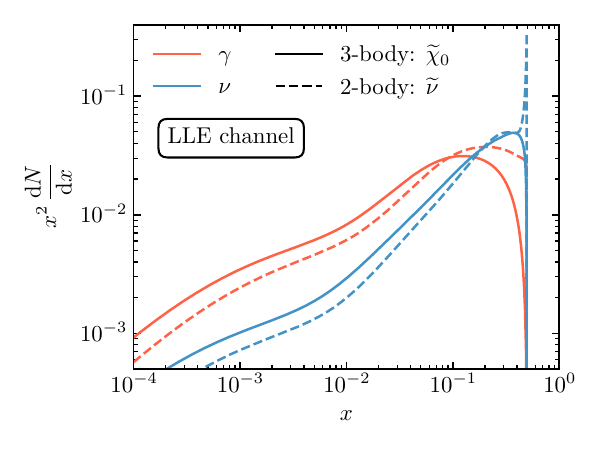}
    \caption{Comparison between the shower spectra in $\gamma$ (red) and $\nu$ (blue, averaged over the three flavours) from sneutrino (dashed lines) and neutralino (solid lines) DM candidates, both with $m_{\text{DM}}=10^{10}~\mathrm{GeV}$ and decaying via the $LLE$ vertex according to two and three-body processes, respectively.}
    \label{fig:comparison_2_3_body_spectra}
\end{figure}

In Fig.~\ref{fig:comparison_2_3_body_spectra} we show a comparison between the shower spectra in photons and neutrinos, coming from the two- and three-body decay spectra for the $LLE$ vertex in the case of a sneutrino and a neutralino DM candidate, respectively, fixing $m_\textup{DM} = 10^{10}~\mathrm{GeV}$. As expected, the three-body decay spectra are shallower and peaked towards lower $x$ with respect to the case of the two-body decay spectra.

\section{Neutrino and gamma-ray bounds for three-body decays}
\label{sec:bounds_indep}

\begin{figure}[t!]
    \centering
    \includegraphics[width=0.49\textwidth]{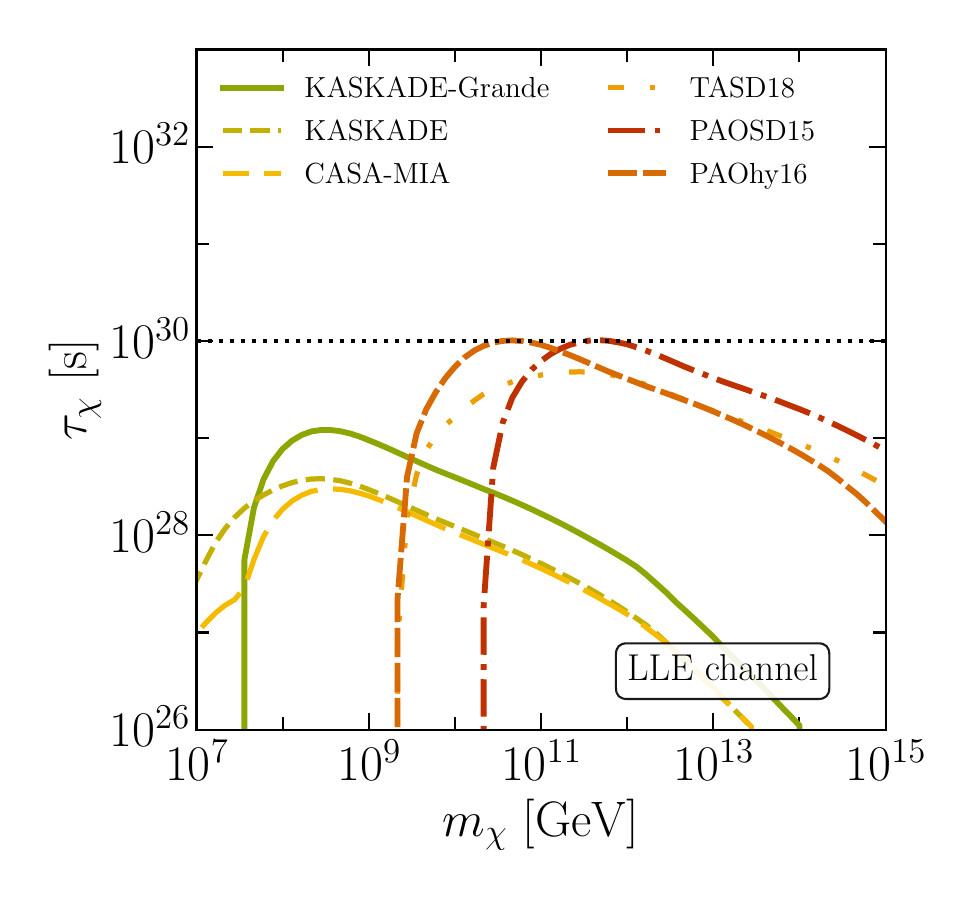}
    \includegraphics[width=0.49\textwidth]{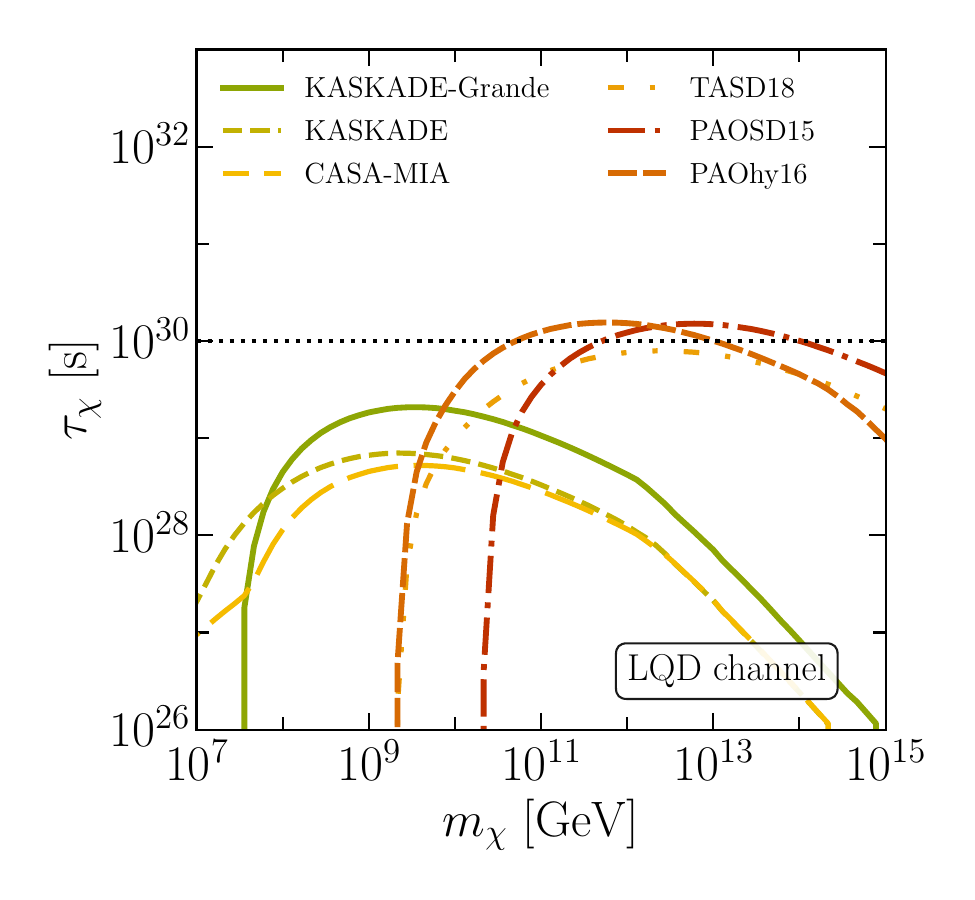}
    \includegraphics[width=0.49\textwidth]{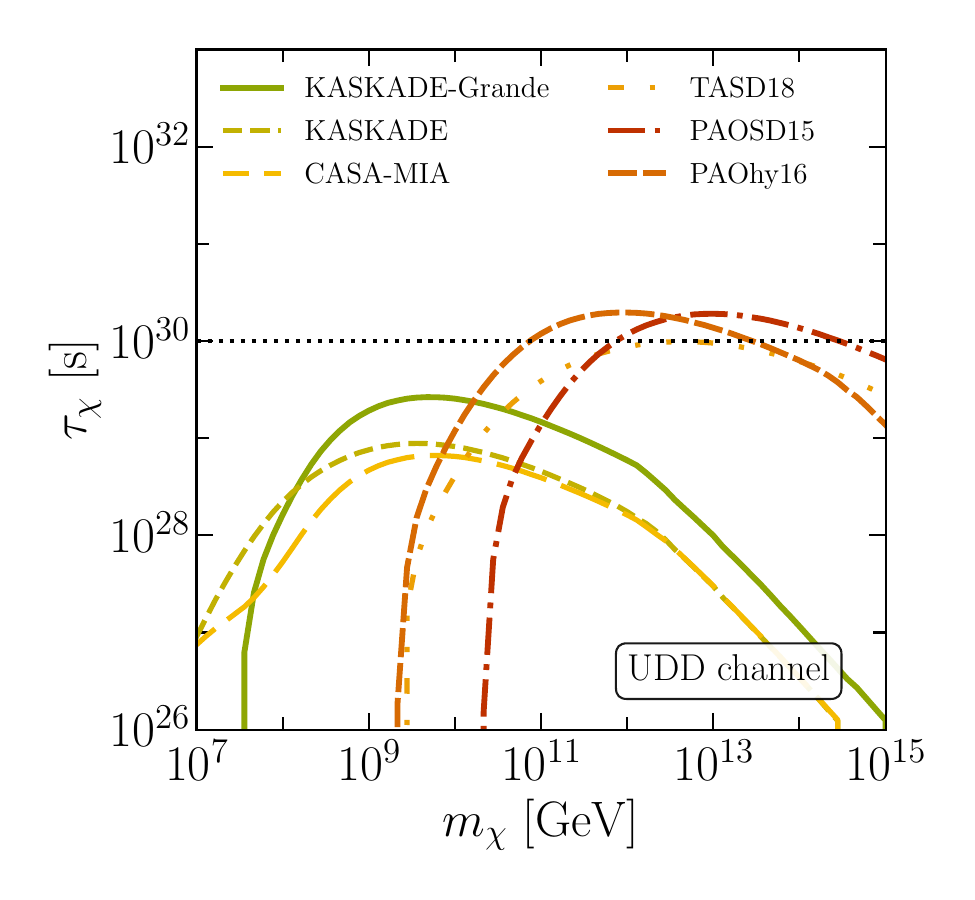}
    \caption{Current gamma-ray limits on the DM lifetime as a function of the DM mass for different experiments (different lines), assuming at a time one of the three different three-body decay channels arising from the RPV couplings (see Tab.~\ref{tab:RPVdecays}): $LLE$ (top left), $LQD$ (top right), and $UDD$ (bottom center). The horizontal dotted line acts as a reference.}
    \label{fig:3body_gamma}
\end{figure}

\begin{figure}[t!]
    \centering
    \includegraphics[width=\textwidth]{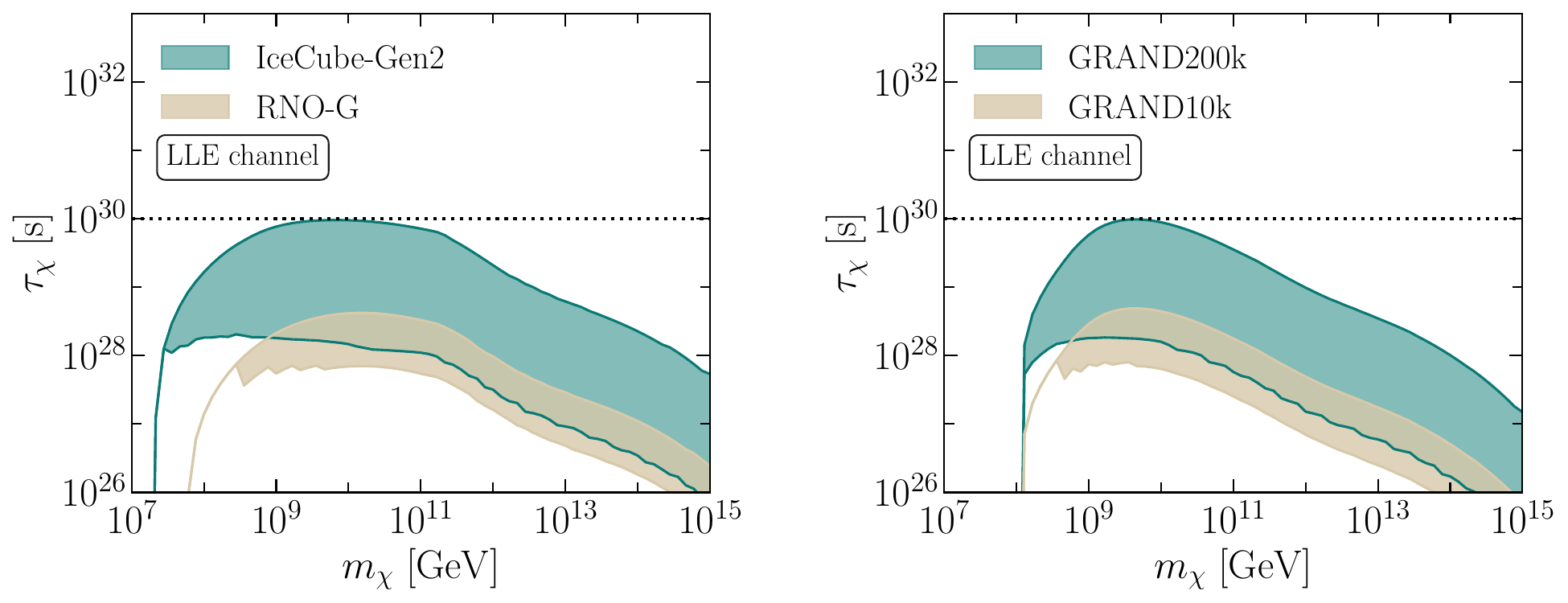}
    \includegraphics[width=\textwidth]{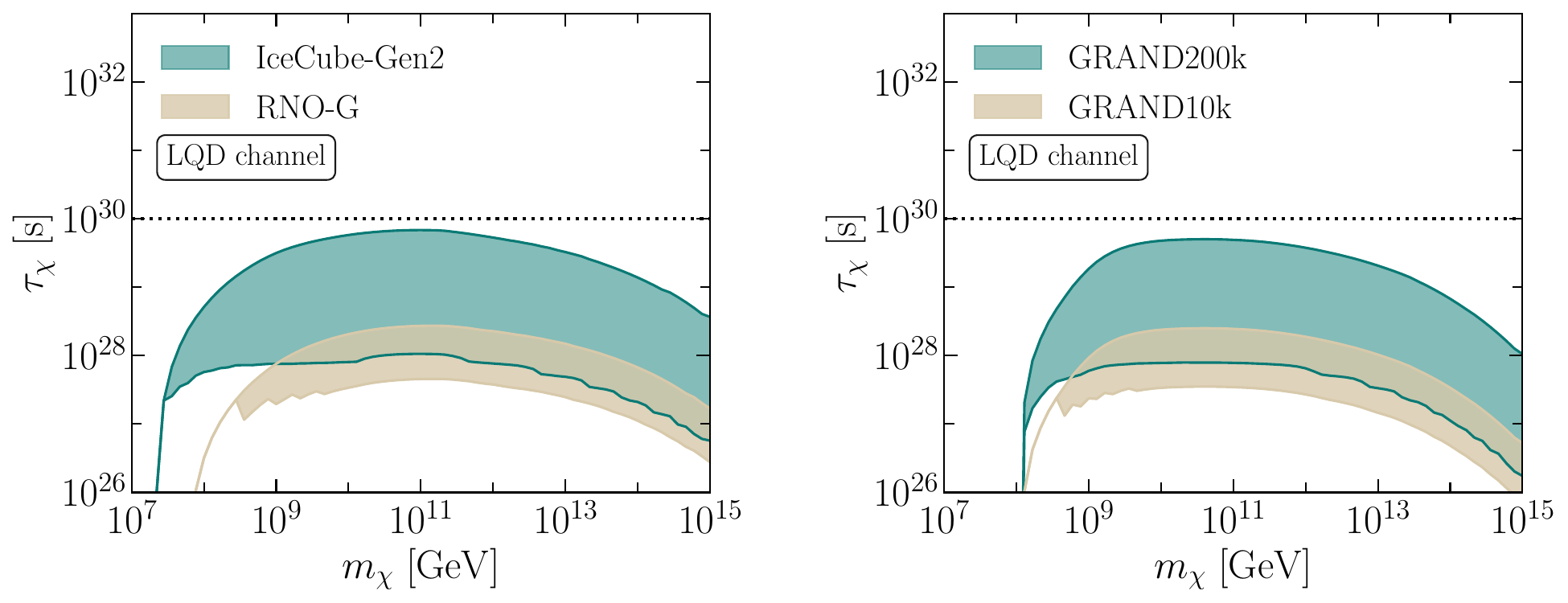}
    \includegraphics[width=\textwidth]{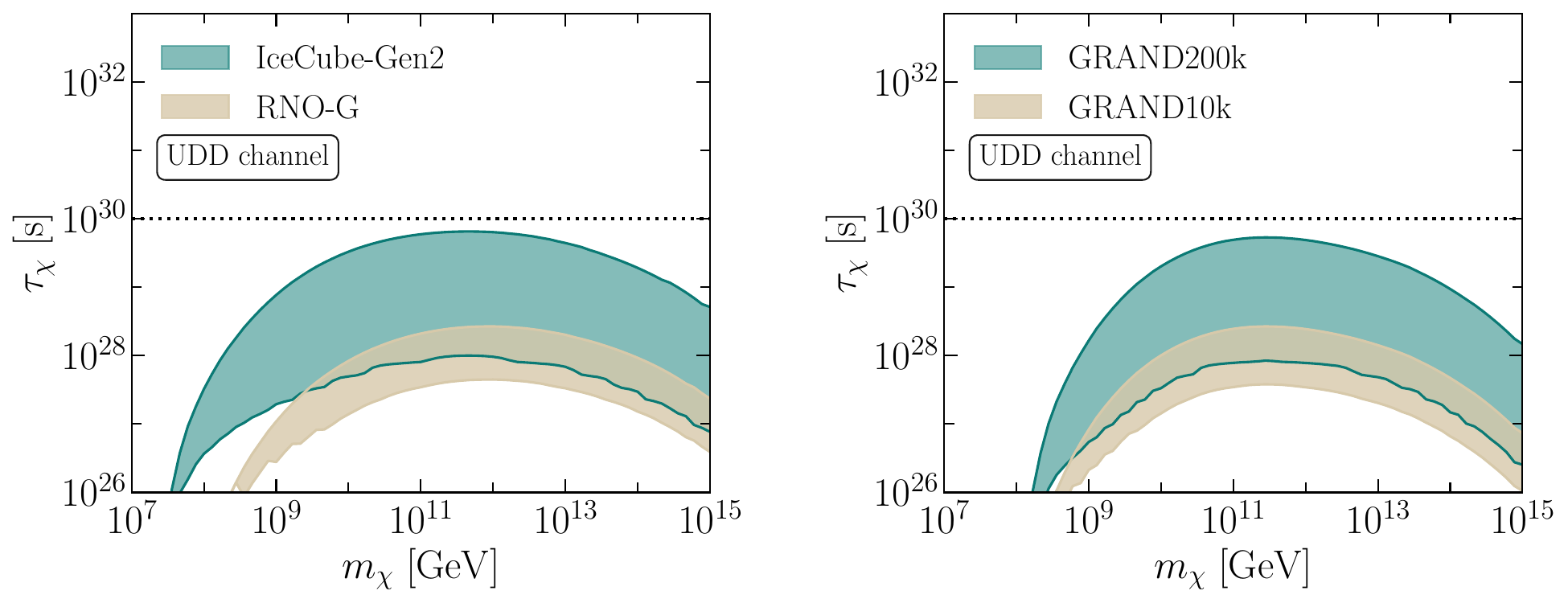}
    \caption{Forecast neutrino limits on the DM lifetime as a function of the DM mass for different future neutrino telescopes, assuming at a time one of the three different three-body decay channels arising from the RPV couplings (see Tab.~\ref{tab:RPVdecays}): $LLE$ (top panels), $LQD$ (center panels), and $UDD$ (bottom panels). The bands cover the uncertainty on the astrophysical neutrino flux which leads to a different expected number of detected neutrino events after 3 years of observation. The horizontal dotted line acts as a reference.}
    \label{fig:3body_nu}
\end{figure}

After describing the statistical procedure in Sec.~\ref{sec:expconstr} and the computation of the energy spectra in Sec.~\ref{sec:spectra}, we can now obtain the lower bounds on the DM lifetime $\tau_\chi$ for DM masses $m_\chi$ in the range from $10^7$ to $10^{15}$~GeV. Here, we focus on
the three-body decay channels reported in Tab.~\ref{tab:RPVdecays}, which occur in the case of neutralino DM. The case of two-body decay channels (sneutrino DM) is discussed in App.~\ref{app:snudecay}. We consider that the decay of DM occurs through only one of the couplings $LLE$, $LQD$ and $UDD$ (see Tab.~\ref{tab:RPVdecays}) at a time. Moreover, we assume that such couplings are independent from the particle flavours. This results in 36, 108, 18 different final states (taking into account different flavour combinations) with equal branching ratio for the $LLE$, $LQD$, $UDD$ scenarios, respectively. Hence, we first discuss the model-independent constraints for each coupling, and then report the constraints projected in the parameter space of the string model featuring superheavy neutralino DM.
\begin{figure}[t!]
    \centering
    \includegraphics[width=0.8\textwidth]{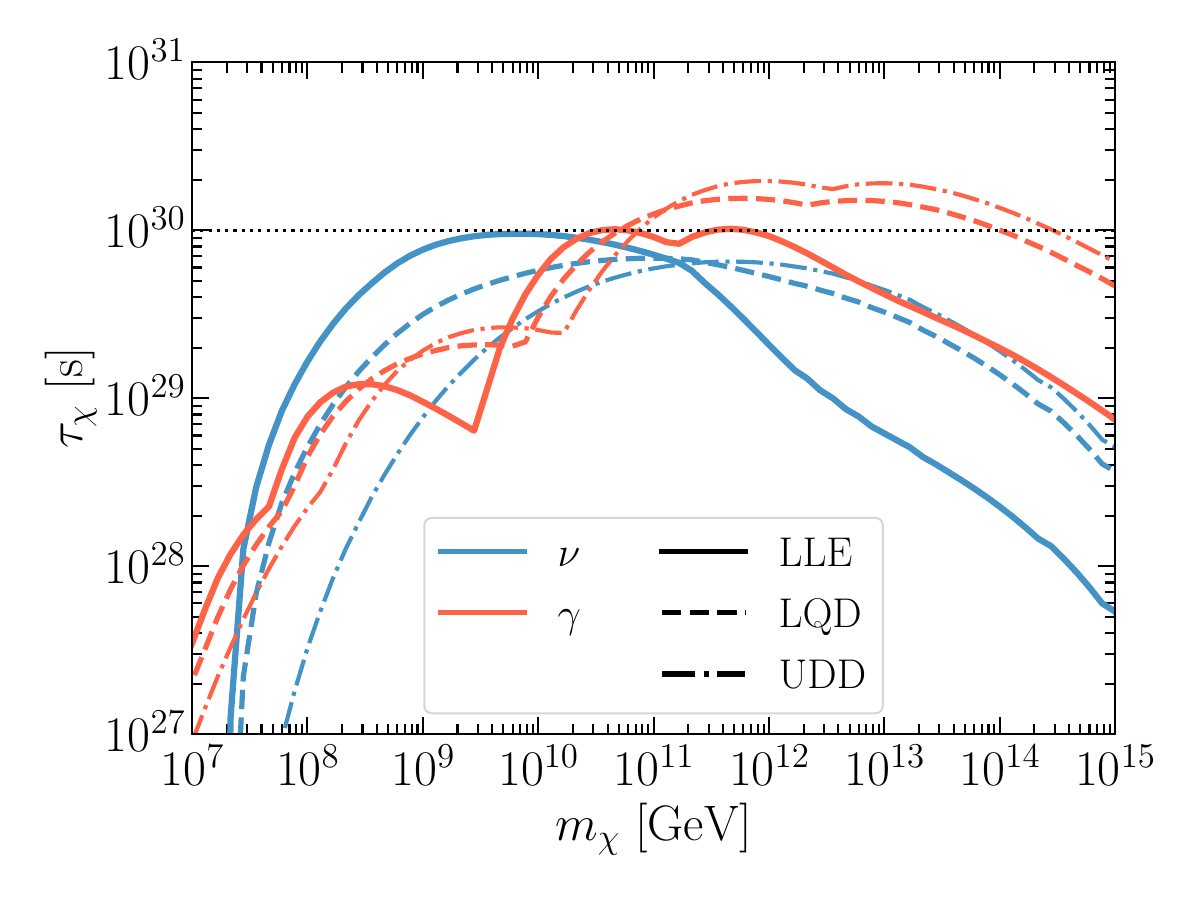}
    \caption{Comparison of the constraints on the DM lifetime placed with all current gamma-ray data (red lines) and with no neutrino events in IceCube-Gen2 after 3 years of data-taking (blue lines). The solid, dashed and dot-dashed lines refer to the $LLE$, $LQD$, $UDD$ channels, respectively (see Tab.~\ref{tab:RPVdecays}).}
    \label{fig:3body_comparison}
\end{figure}

In Figs.~\ref{fig:3body_gamma} and~\ref{fig:3body_nu} we report the bounds obtained with current gamma-ray data and with forecast neutrino data, respectively, in the case of the three different decay channels. In the former figure, different lines correspond to the bounds placed by different telescopes. In the latter, the left (right) panels show the bounds for the future neutrino telescopes RNO-G and IceCube-Gen2 (GRAND10k and GRAND200k). Moreover, the neutrino forecast bounds are shown as bands which correspond to the uncertainty on the different scenarios for the expected neutrino observations after 3 years of data-taking according to the astrophysical pulsar and cosmogenic neutrinos, as well as to zero neutrino events (null observation of neutrino events). The latter provides the strongest constraints on the DM properties from neutrino data. The common behaviour of all the constraints is due to the fact that each experiment is sensitive to a specific energy range. This translates into a specific interval of the DM mass that can be probed.

In Fig.~\ref{fig:3body_comparison} we make a direct comparison between the constraints placed by all the gamma-ray experiments (red lines) and the ones which will be placed by IceCube-Gen2 assuming zero neutrino events (blue lines). Different line styles correspond to different DM decay channels. 
As can been seen, we generally expect future neutrino data to provide stronger bounds on the DM lifetime for DM masses below $10^{11}$ GeV. This behaviour depends on the decay channels. In particular, for the pure leptophilic $LLE$ channel, which involves only leptons (solid lines), we expect an improvement on the constraints from the neutrino data for a larger range of DM masses. On the other hand, the impact of neutrino data is expected to be very limited in the case of the pure hadronic $UDD$ channel (dot-dashed lines). This is simply due to the different energy injection into UHE photons and neutrinos for the different decay channels. We emphasise that our forecast analysis is not strongly sensitive to spectral features and to angular DM anisotropy since it is based on energy-integrated and angle-averaged quantities. However, the neutrino constraints could be further improved with a more refined analysis once actual data will be taken by the future telescopes.

\begin{figure}[t!]
    \centering
    \includegraphics[width=0.49\textwidth]{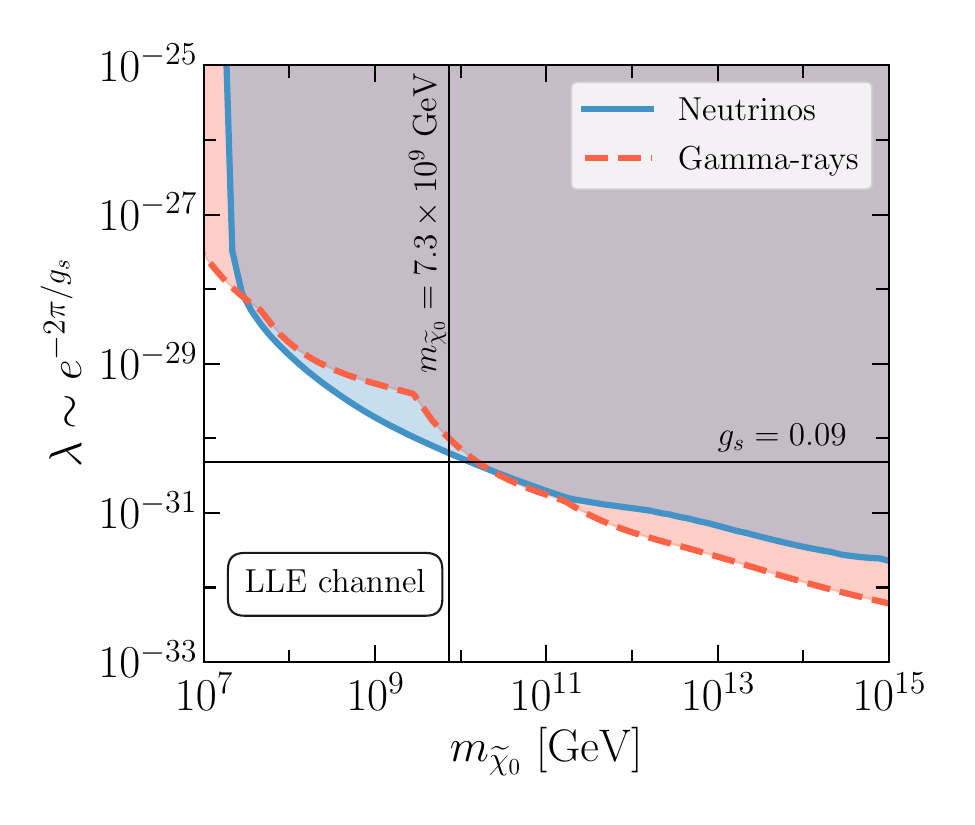}
    \includegraphics[width=0.49\textwidth]{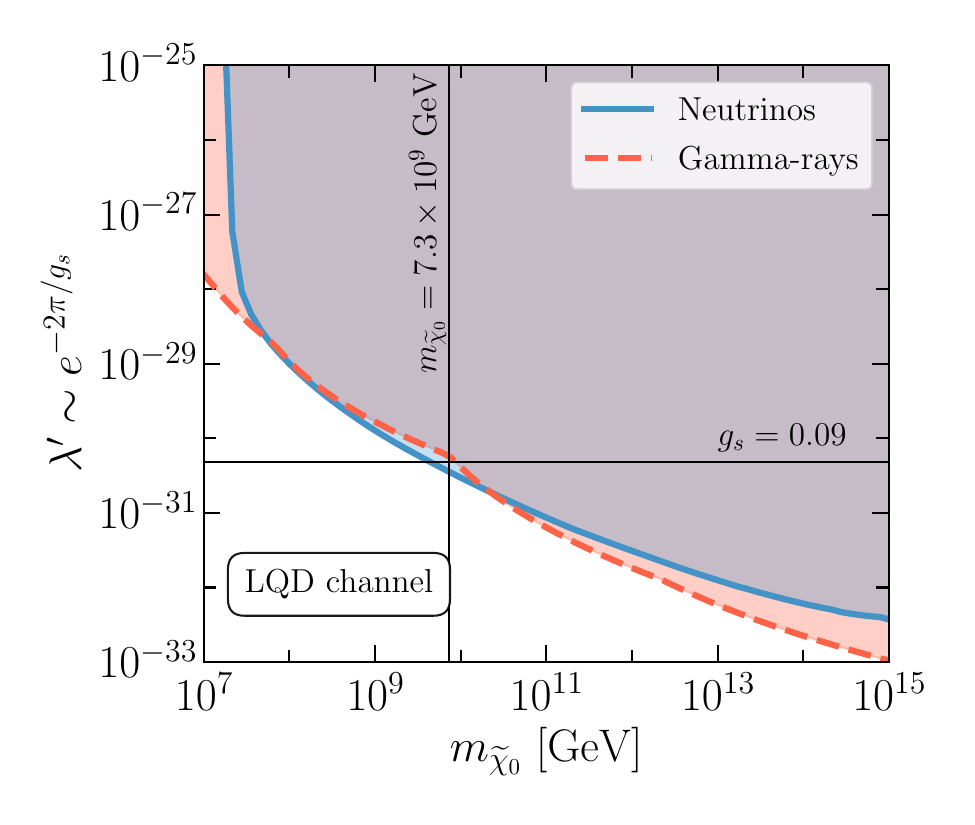}
    \includegraphics[width=0.49\textwidth]{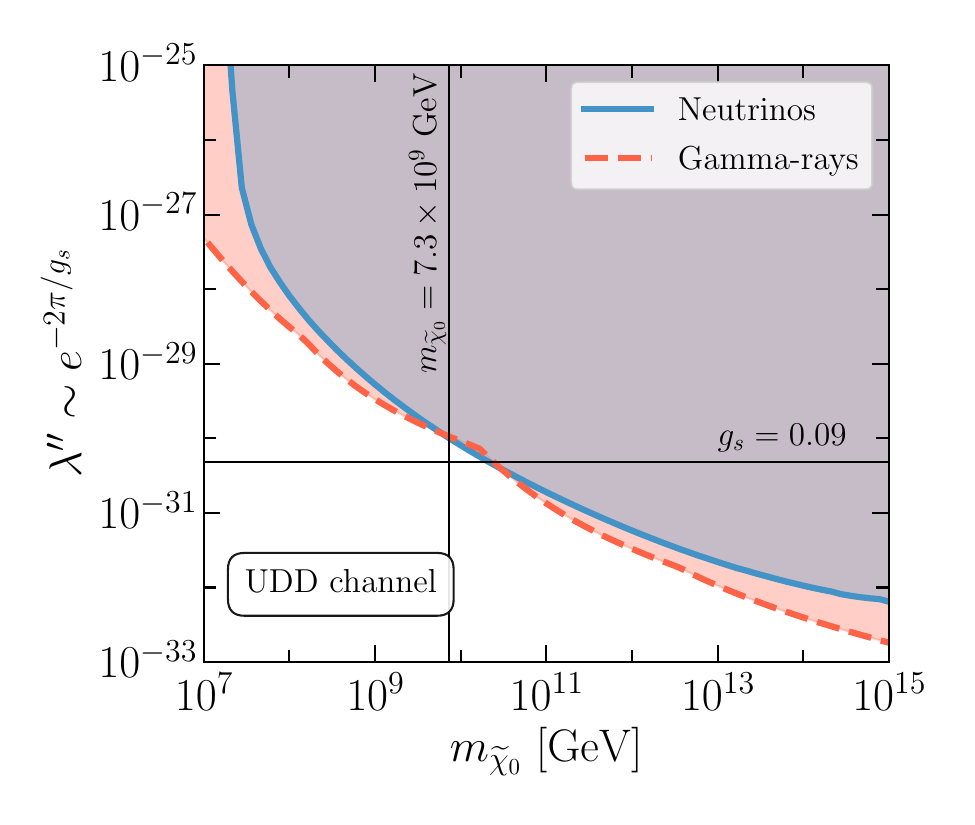}
    \caption{Upper bounds on the $LLE$ (top left), $LQD$ (top right) and $UDD$ (bottom center) couplings from current gamma-ray (red dashed lines) and future neutrino (blue solid lines) data, as a function of the neutralino DM mass $m_{\widetilde{\chi}_0}$. For the three cases, the mediator mass is fixed according to the RGE running (see the main text). The thin black lines display a benchmark value for the string coupling $g_s=0.09$ and the model prediction for $m_{\widetilde{\chi}_0}$.}
    \label{fig:lambda}
\end{figure}

\begin{figure}[t!]
    \centering
    \includegraphics[width=0.49\textwidth]{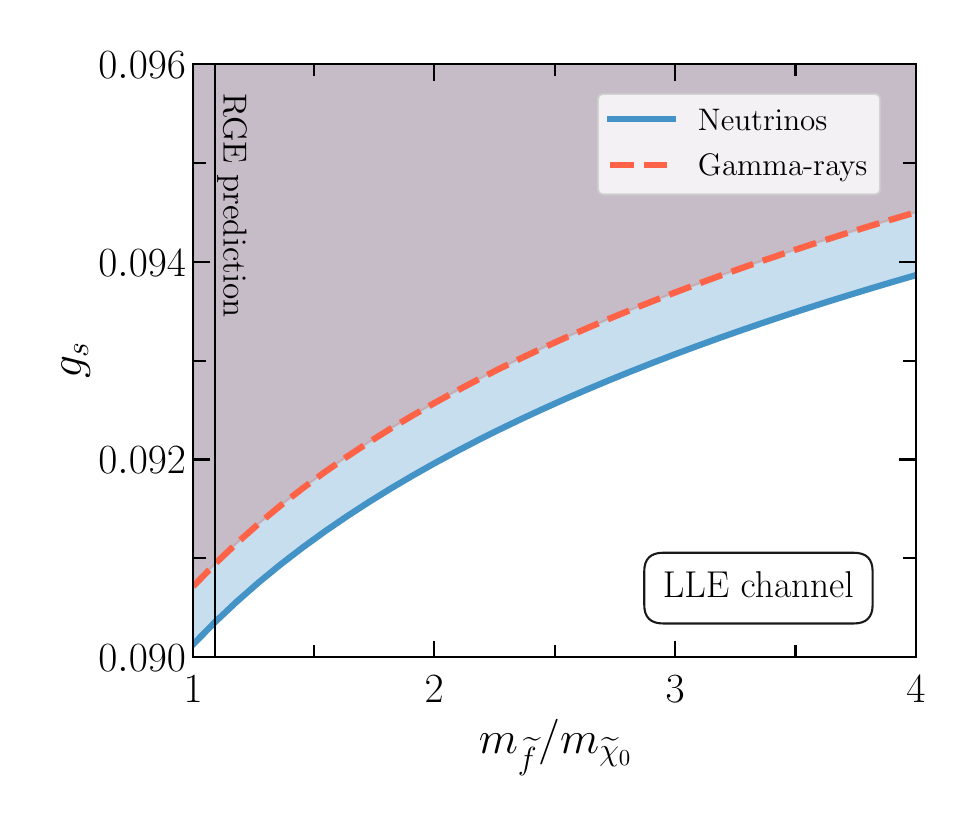}
    \includegraphics[width=0.49\textwidth]{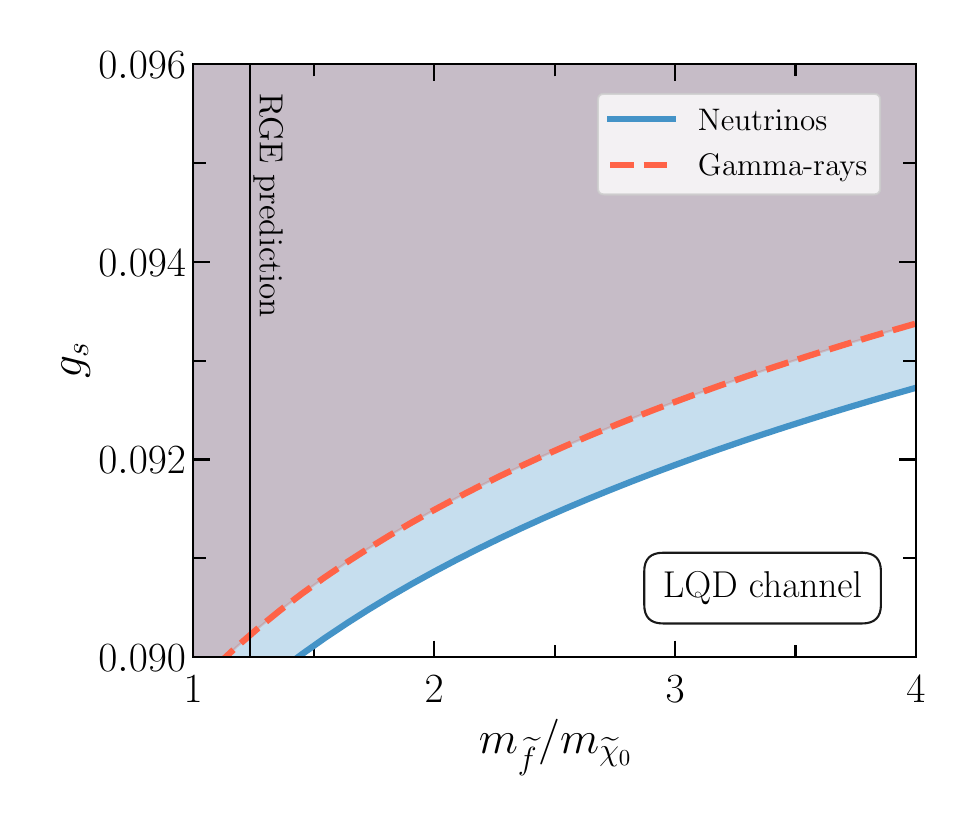}
    \includegraphics[width=0.49\textwidth]{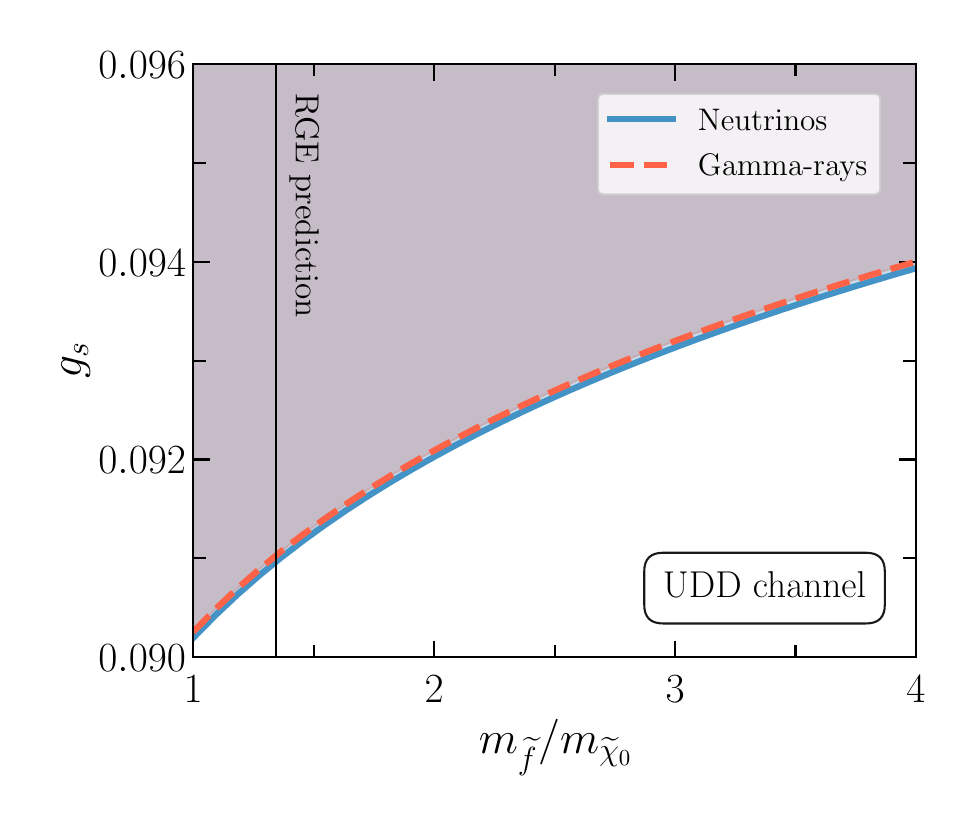}
    \caption{Upper bounds on the string coupling $g_s$ as a function of the mass ratio $m_{\widetilde{f}} / m_{\widetilde{\chi}_0}$ with $m_{\widetilde{\chi}_0}=7.3 \times 10^9~{\rm GeV}$. The different panels correspond to the different RPV couplings of the model: $LLE$ (top left), $LQD$ (top right) and $UDD$ (bottom center). Red dashed lines (blue solid lines) refer to the current gamma-ray (future neutrino) limits. The vertical thin black lines display the model prediction for the mass ratio.}
    \label{fig:gs}
\end{figure}

All of the previous bounds hold for a generic DM candidate with mass $m_\chi$ and three-body decay channels summarised in Tab.~\ref{tab:RPVdecays}. For the specific model discussed in Sec.~\ref{sec:model}, we can recast the constraints on the DM lifetime $\tau_\chi$ into model-dependent bounds on the RPV coupling $\lambda$ by means of Eqs.~\eqref{eq:gamma1_neutralino},~\eqref{eq:gamma2_neutralino} and~\eqref{eq:gamma3_neutralino} and, remarkably, on the string coupling $g_s$ through the relation $\lambda \sim e^{-2\pi/g_s}$. To obtain such bounds, indeed one needs to also define the mass of the mediator (see Fig.~\ref{fig:neutralino_rpv}) as well as the RGE running of the fine-structure constant which has to be computed at an energy scale $Q \sim 10^{10}~{\rm GeV}$. As shown in Fig.~\ref{fig:RGE}, at this scale the mass of the neutralino DM is $m_{\widetilde{\chi}_0}=7.3 \times 10^{9}~{\rm GeV}$, while the averaged masses $m_{\widetilde{f}}$ of the mediators are $8.0 \times 10^{9}~{\rm GeV}$, $9.1 \times 10^{9}~{\rm GeV}$ and $9.9 \times 10^{9}~{\rm GeV}$, for the three different couplings $LLE$, $LQD$ and $UDD$, respectively. Moreover, the fine-structure constant is $\alpha (m^2_{\widetilde{\chi}_0}) =0.01172$.

In Fig.~\ref{fig:lambda} we report the constraints on the three different couplings as a function of the neutralino DM mass $m_{\widetilde{\chi}_0}$. The red dashed lines correspond to the gamma-ray constraints, while the blue solid ones to the strongest bounds that can be placed with future neutrino observations assuming zero neutrino events (no astrophysical backgrounds). The vertical thin black lines mark the prediction of the model for the DM mass, while the horizonal lines show a benchmark value for the string coupling $g_s$. Hence, these bounds can be then translated into constraints on $g_s$ as done in Fig.~\ref{fig:gs} where we fix the DM mass to be $m_{\widetilde{\chi}_0}=7.3 \times 10^{9}~{\rm GeV}$ and vary the mediator masses. Here, the vertical lines correspond to the mediator masses obtained from the RGE running. 

Interestingly, for the DM mass predicted by the model, the allowed values for $g_s$ are $g_s \lesssim 0.1$. This implies that the phenomenological constraints from UHE gamma rays and neutrino observations coincide with the theoretical requirement on the string coupling to be in the regime where the low-energy effective field theory is under perturbative control. Let us also mention that in string compactifications the value of $g_s$ is determined dynamically as the vacuum expectation value of the dilaton field which is stabilised by background 3-form fluxes. Given that different possible choices of these flux quanta are compatible with tadpole cancellation, a landscape of string vacua arises with an approximately uniform distribution of the string coupling, i.e. $N_{\text{vac}}(g_s)\simeq g_s$ \cite{Broeckel:2020fdz, Cicoli:2022chj, Ashok:2003gk, Denef:2004ze}. One might therefore naively think that the phenomenological constraint $\lambda \lesssim 10^{-30}$ from Fig. \ref{fig:lambda} could leave only a very sparse region of the landscape. This is however not true since the RPV coupling $\lambda$ depends exponentially on $g_s$, $\lambda \sim e^{-2\pi/g_s}$, and so 
the distribution of $\lambda$ turns out to be milder than logarithmic:
\begin{equation}
d\lambda = \frac{\partial \lambda}{\partial g_s}\,d g_s \simeq  \frac{1}{2\pi} \left(\ln\lambda\right)^2 \lambda \, d N_{\text{vac}}
\qquad\Leftrightarrow\qquad N_{\text{vac}}(\lambda)\sim - \left(\ln\lambda\right)^{-1} = o(\ln\lambda)\,.
\end{equation}
Thus the increase of the number of vacua for larger values of $\lambda$ is milder than logarithmic, implying that the region with $\lambda \lesssim 10^{-30}$ contains still several string vacua which are in agreement with current data and can be probed with future DM observations.

\section{Conclusions}
\label{sec:concl}

In this paper we have studied indirect detection signals from decaying superheavy DM. Generic features of string constructions (namely a high scale of SUSY breaking and epochs of EMD driven by string moduli) provide a suitable framework to accommodate superheavy LSP DM and obtain the correct DM relic abundance \cite{Allahverdi:2020uax}. Moreover, tiny RPV couplings induced by stringy instantons naturally lead to a DM instability with decay lifetimes longer than the age of the Universe.   

As a case study, we considered an explicit string model with high scale SUSY breaking that gives rise to successful inflation and yields the correct DM relic abundance in the $10^{10}-10^{11}$ GeV mass range. We derived the mass spectrum of SUSY particles for a benchmark point that realises the MSSM at low energies and found that the LSP DM is a Bino-like neutralino. The RPV couplings lead to three-body decays of DM to quarks and leptons.

By adopting a model-independent approach, we computed the resulting gamma-ray and neutrino fluxes as a function of the neutralino mass and lifetime. We obtained constraints in the mass-lifetime plane by imposing the limits from current experiments and using the sensitivities of future neutrino observations. The most stringent limits on the lifetime, $\tau_\chi \gtrsim 10^{30}$ s happen for $m_\chi\sim 10^9-10^{11}$ GeV. We found that these bounds have a rather mild dependence on the specific decay channel (leptonic/hadronic/mixed as well as flavour combination of final states). Furthermore, they do not change significantly for Higgsino-like or Wino-like neutralino DM. It is interesting to note that for the specific string model considered, these limits result in an upper bound of ${\cal O}(0.1)$ on the string coupling which is compatible with the effective field theory being in the controlled regime where perturbation theory does not break down.

Our work provides an explicit example for indirect detection of superheavy DM which would otherwise evade (in)direct detection and collider searches, in the context of a UV-complete model from type IIb string compactifications. A natural extension of this model could include RH neutrinos (and their superpartners) that are usually introduced to address neutrino mass and mixing. This results in new final states for DM decay that include the RH neutrinos. This is worth exploring as such high energy RH neutrinos might provide an explanation for the recent anomalous events reported by IceCube and ANITA experiments (for example, see~\cite{HMP}).   

\section*{Acknowledgments}

We thank Nicolas Rodd for discussions on the energy spectra and Gwenha\"el De Wasseige for discussions about neutrino searches.
Rouzbeh Allahverdi wishes to thank the Department of Theoretical Physics at CERN for their kind hospitality while this work was being completed. The work of Rouzbeh Allahverdi is supported in part by NSF Grant No. PHY-2210367.
Fabio Maltoni and Chiara Arina acknowledge the support by the F.R.S.-FNRS under the “Excellence of Science” EOS be.h project no. 30820817.
The work of Marco Chianese is supported by the research grant number 2022E2J4RK ``PANTHEON: Perspectives in Astroparticle and Neutrino THEory with Old and New messengers'' under the program PRIN 2022 funded by the Italian Ministero dell'Universit\`a e della Ricerca (MUR) and by the research project TAsP (Theoretical Astroparticle Physics) funded by the Istituto Nazionale di Fisica Nucleare (INFN). Jacek K. Osi\'nski is supported by the grant ``AstroCeNT: Particle Astrophysics Science and Technology Centre" carried out within the International Research Agendas programme of the Foundation for Polish Science financed by the European Union under the European Regional Development Fund. The work of Michele Cicoli contributes to the COST Action COSMIC WISPers CA21106,
supported by COST (European Cooperation in Science and Technology)".

\appendix

\section{SUSY mass spectrum and RGE running}\label{app:RGE}

The one-loop RGEs for the gauge couplings, $g_a$, and gaugino masses, $M_a$, are given by:
\begin{eqnarray}\label{eq:RGE1}
    \frac{{\rm d} g_a}{{\rm d}t} & = & \frac{g_a^3}{16 \pi^2} b_a\,,\\
    \frac{{\rm d} M_a}{{\rm d}t} & = & \frac{2 g_a^2}{16 \pi^2} b_a M_a\,,
    \label{eq:RGE2}
\end{eqnarray}
with $b_a = 33/5,\, 1,\, -3$ for $a = 1,2,3$ in the MSSM~\cite{Martin:1997ns}. 
To compute the scalar masses at low energies we make use of the approximate analytical solutions for the running of the squared masses~\cite{Martin:1997ns,Conlon:2007xv}:  
\begin{eqnarray}\label{eq:RGEscalar1}
    m_{\tilde{d}_L}^2(Q) & = & m_{\tilde{d}_L}^2(Q_0) + K_3(Q) + K_2(Q) + \frac{1}{36}K_1(Q) + \Delta_{\tilde{d}_L}\,,\\
    m_{\tilde{u}_L}^2(Q) & = & m_{\tilde{u}_L}^2(Q_0) + K_3(Q) + K_2(Q) + \frac{1}{36}K_1(Q) + \Delta_{\tilde{u}_L}\,,\\
    m_{\tilde{d}_R}^2(Q) & = & m_{\tilde{d}_R}^2(Q_0) + K_3(Q) + \frac{1}{9}K_1(Q) + \Delta_{\tilde{d}_R}\,,\\
    m_{\tilde{u}_R}^2(Q) & = & m_{\tilde{u}_R}^2(Q_0) + K_3(Q) + \frac{4}{9}K_1(Q) + \Delta_{\tilde{u}_R}\,,\\
    m_{\tilde{e}_L}^2(Q) & = & m_{\tilde{e}_L}^2(Q_0) + K_2(Q) + \frac{1}{4}K_1(Q) + \Delta_{\tilde{e}_L}\,,\\
    m_{\tilde{e}_R}^2(Q) & = & m_{\tilde{e}_R}^2(Q_0) + K_1(Q) + \Delta_{\tilde{e}_R}\,,\\
    m_{\tilde{\nu}}^2(Q) & = & m_{\tilde{\nu}}^2(Q_0) + K_2(Q) + \frac{1}{4}K_1(Q) + \Delta_{\tilde{\nu}}\,,
    \label{eq:RGEscalar2}
\end{eqnarray}
where $Q_0$ is the high scale at which the boundary conditions are evaluated, while $Q$ is the low scale of interest. In our estimation of the low-energy sparticle spectrum, we neglect the contributions from the splittings $\Delta$, which originate from the D-terms, as they are typically smaller than the other contributions. The functions $K_a(Q)$ are given, at one-loop order, by:
\begin{equation}
    K_a(Q) = \{3/5,\, 3/4,\, 4/3\}\, \times \frac{1}{2\pi^2} \int_{\ln{Q}}^{\ln{Q_0}}{\rm d}t\, g_a^2(t)\, |M_a(t)|^2\qquad (a = 1,2,3)\,,
\end{equation}
where $g_a(Q)$ and $M_a(Q)$ are obtained from solutions to Eqs.~\eqref{eq:RGE1} and \eqref{eq:RGE2}. 

\section{Sneutrino dark matter and two-body decay spectra}\label{app:snudecay}

In the MSSM typically the sneutrino cannot be the LSP, as one can see from Fig.~\ref{fig:RGE}, unless the GUT unification condition of scalar and gaugino masses is uplifted. A possibility to have a sneutrino LSP, and hence a good DM candidate, is to extend the MSSM content with a right-handed neutrino superfield. The right-handed sneutrino can then be DM (with or without mixing with the left-handed sneutrino). In this way it is possible to tackle at the same time the neutrino mass issue as well (for example, see~\cite{Matchev,Allahverdi:2007wt, Arina:2007tm, Arina:2008bb, Munoz, Allahverdi:2009ae, Allahverdi:2009se} for further discussions on how to achieve sneutrino DM).

\begin{figure}[t!]
    \centering
    \includegraphics{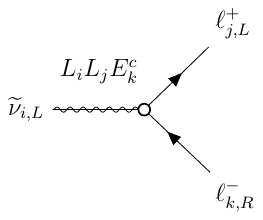}
    \includegraphics{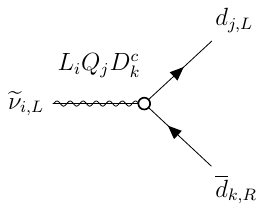}
    \caption{Feynman diagrams illustrating the decay of the LSP $\widetilde{\nu}_\ell$ into two SM fermions.
    The RPV vertex is highlighted by the white dots.}
    \label{fig:sneutrino_rpv}
\end{figure}

Sneutrino DM is made directly unstable by the RPV terms in \eqref{eq:explicit}. However, note that only the $LLE$ and $LQD$ terms can contribute to $\widetilde{\nu}$ decay, because of the superfield $L$.
The sneutrino decays into two fermions and the main decay diagrams are illustrated in Fig.~\ref{fig:sneutrino_rpv}.
The expressions for the sneutrino decay into fermions are given by~\cite{Barger:1995fv}:
\begin{eqnarray}
    \Gamma_{\widetilde{\nu}_i}^{LLE} & = & \frac{1}{16 \pi} \lambda^2_{ijk} m_{\widetilde{\nu}_i}\,,\\
    \Gamma_{\widetilde{\nu}_i}^{LQD} & = & \frac{3}{16 \pi} \lambda'^{2}_{ijk} m_{\widetilde{\nu}_i}\,,
\end{eqnarray}
where the fermion masses have been neglected.

\begin{table}[tp]
\caption{Two-body sneutrino final states induced by the RPV vertices in Eqs.~\eqref{eq:explicit}. As usual, $i,j,k$ denote the generation of fermions.}
\centering
\begin{tabular}{ccc}
\toprule
 & \multicolumn{2}{c}{Couplings / channels} \\
 &~~$\lambda_{ijk}$ / LLE channel~~&~~$\lambda_{ijk}^{\prime}$ / LQD channel~~\\
\midrule
$\widetilde{\nu}_{i,L}$ & $\ell_{j,L}^+ \ell_{k,R}^-$ & $\overline{d}_{j,L} d_{k,R}$  \\
$\widetilde{\nu}_{j,L}$ & $\ell_{i,L}^+ \ell_{k,R}^-$ &  \\
\bottomrule
\end{tabular}
\label{tab:sneutrino_RPVdecays}
\end{table}

The possible final states are given in Tab.~\ref{tab:sneutrino_RPVdecays}. It is interesting to remark that it is not possible to have monochromatic neutrino lines from the two-body decay, because of the structure of the RPV terms. This is in contrast with the case of annihilating sneutrino DM (for example, see~\cite{Arina:2015zoa}). 

From the numerical point of view, the sneutrino DM decay is simpler to handle than the one of $\widetilde{\chi}_0$, to obtain the energy spectra. The sneutrino decays via $LLE$ into two charged leptons of different flavour, and there are 12 possible contributions, such as $\widetilde{\nu} \to e^+ \mu^-, \, \widetilde{\nu} \to \tau^+ \mu^-,\, ...$. The $LQD$ vertex gives rise to only 6 final states featuring one left-handed quark and one right-handed down quark, because it contains only one $L$, such as $\widetilde{\nu} \to \overline{d}\, d,\, \widetilde{\chi}_0 \to \overline{s}\, b,\, ...$.
To simulate the sneutrino decay we used the tool \hdmspectra, which allows one to obtain the spectra for two-body DM decay.

\begin{figure}[t!]
    \centering
    \includegraphics[width=\textwidth]{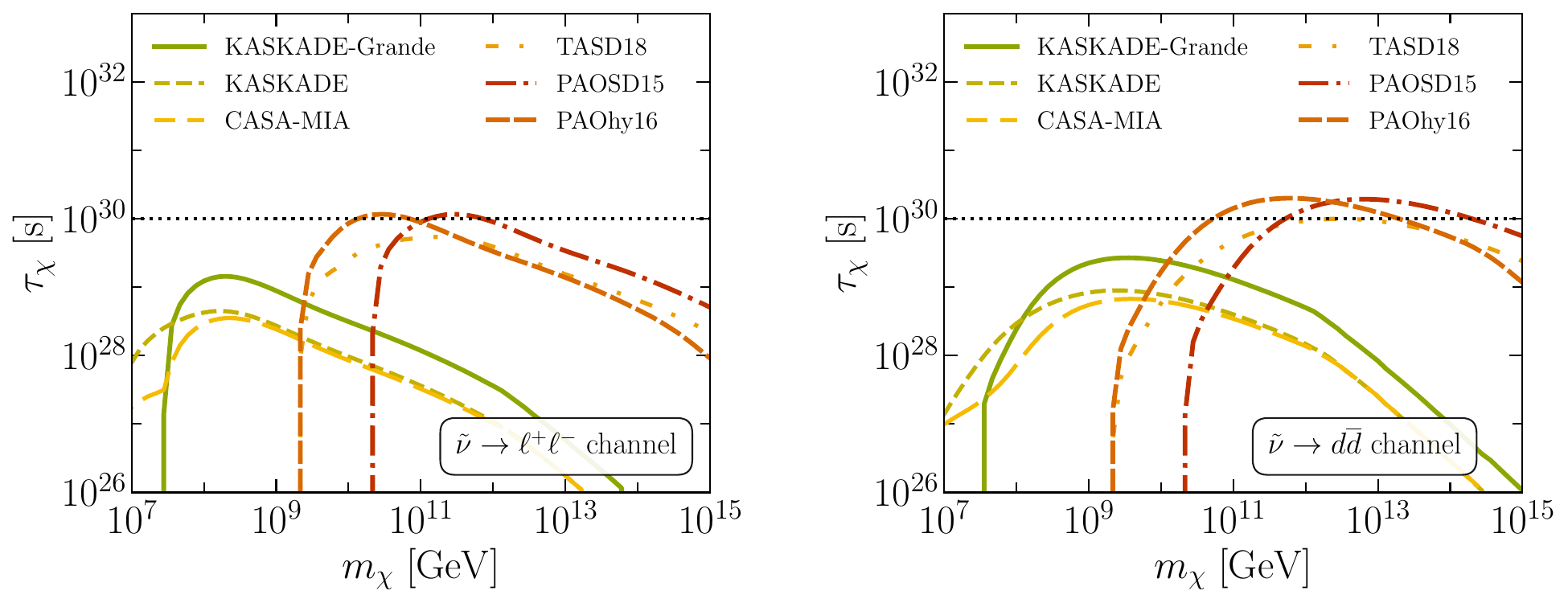}
    \includegraphics[width=\textwidth]{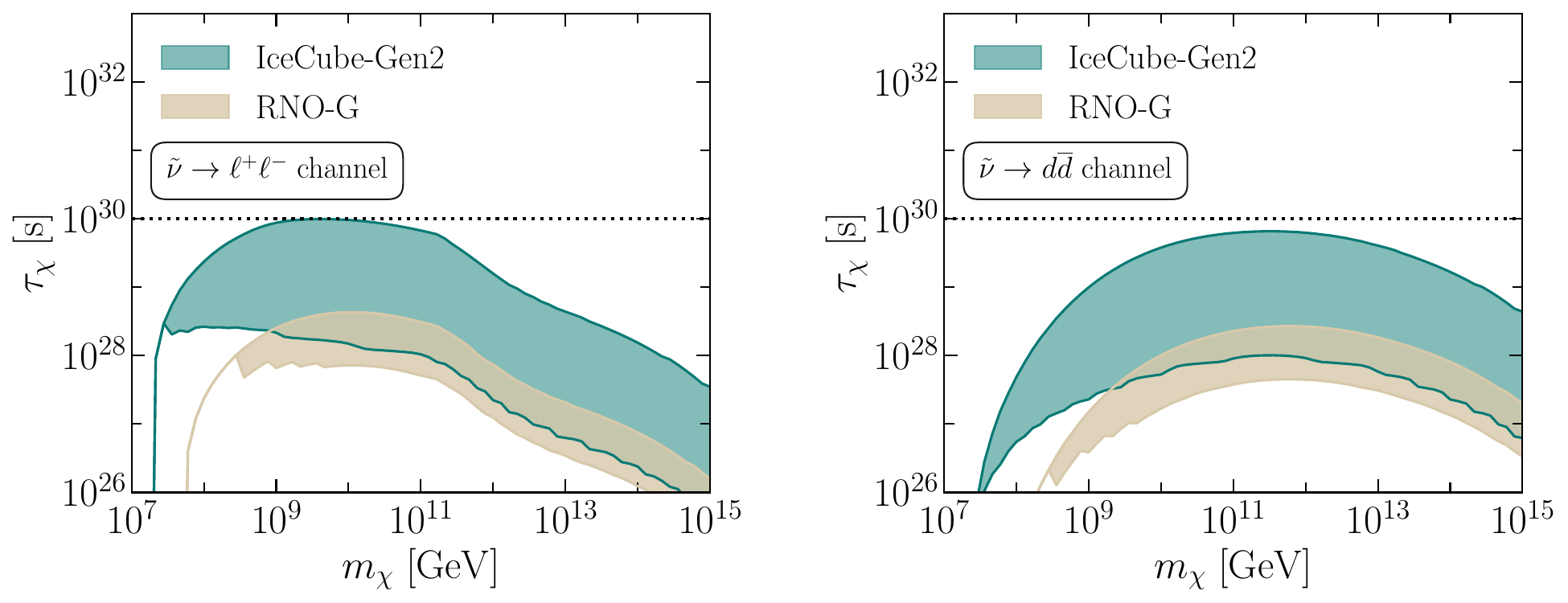}
    \includegraphics[width=\textwidth]{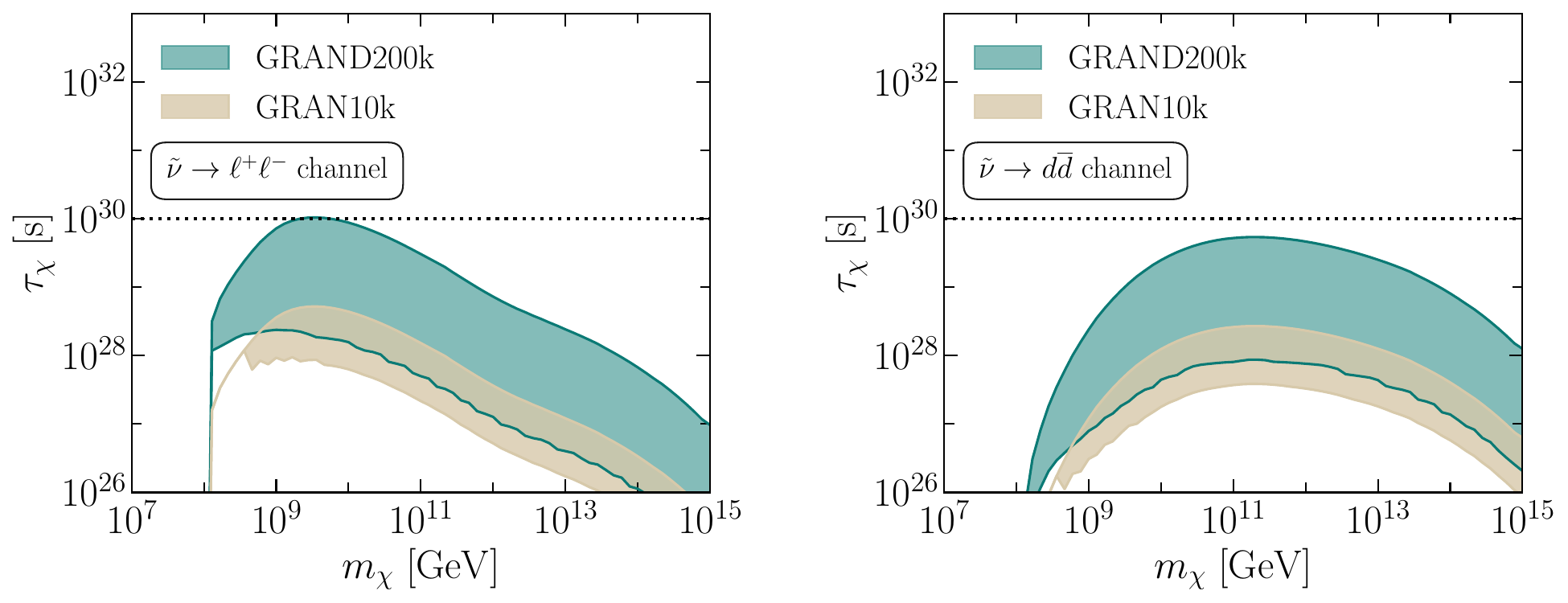}
    \caption{Current gamma-ray limits (top panels) and future neutrino limits (center and bottom panels) on the DM lifetime as a function of the DM mass in case of sneutrino DM $\widetilde{\nu}$ which can decay into two particles as $\widetilde{\nu} \rightarrow \ell^+ \ell^-$ (left panels) and $\widetilde{\nu} \rightarrow d \overline{d}$ (right panels). A sum over different flavour combinations of the final states is performed. The neutrino bands cover the uncertainty on the astrophysical UHE neutrino background for DM searches.}
    \label{fig:2body}
\end{figure}

In Fig.~\ref{fig:2body} we show the lifetime constraints on the basis of gamma-ray data and future neutrino predictions, for a sneutrino DM candidate, in the case of the two RPV decay channels.

The exclusion bounds from gamma rays and the sensitivity reach of neutrino telescopes are rather similar to the case of neutralino DM, as the information coming from the energy spectra is integrated over the full interval of sensitivity of the experiment. This has the effect of making less prominent the differences between two-body and three-body decay spectra.

\bibliographystyle{JHEP}
\bibliography{biblio}

\end{document}